\documentclass[10pt,aps,prd,superscriptaddress,showpacs,twocolumn,nofootinbib,floatfix,amsmath,amssymb]{revtex4-1}
\usepackage{mathtools}
\usepackage[usenames,dvipsnames]{xcolor}
\usepackage{color}
\usepackage{braket}
\usepackage{bm}
\usepackage{graphicx}
\usepackage{hyperref}

\renewcommand{\vec}[1]{\bm{\mathrm{#1}}}
\def\mat#1{\bm{\mathrm{#1}}}
\def\op#1{\hat{#1}}
\def\opvec#1{\op{\vec{#1}}}

\def\abs#1{\left\lvert{#1}\right\rvert}

\begin{document}

\title{Sustainable entanglement production from a quantum field}
\author{Eduardo Mart\'{i}n-Mart\'{i}nez}
\affiliation{Institute for Quantum Computing, University of Waterloo, Waterloo, Ontario, N2L 3G1, Canada}
\affiliation{Dept. Applied Math., University of Waterloo, 200 University
Av W, Waterloo, Ontario, N2L 3G1, Canada}
\affiliation{Perimeter Institute for Theoretical Physics, 31 Caroline St N, Waterloo, ON, N2L 2Y5, Canada}
\author{Eric G. Brown}
\affiliation{Department of Physics and Astronomy, University of Waterloo, Waterloo, Ontario N2L 3G1, Canada}
\author{William Donnelly}
\affiliation{Dept. Applied Math., University of Waterloo, 200 University
Av W, Waterloo, Ontario, N2L 3G1, Canada}
\author{Achim Kempf}
\affiliation{Institute for Quantum Computing, University of Waterloo, Waterloo, Ontario, N2L 3G1, Canada}
\affiliation{Dept. Applied Math., University of Waterloo, 200 University
Av W, Waterloo, Ontario, N2L 3G1, Canada}
\affiliation{Perimeter Institute for Theoretical Physics, 31 Caroline St N, Waterloo, ON, N2L 2Y5, Canada}
\affiliation{Department of Physics and Astronomy, University of Waterloo, Waterloo, Ontario N2L 3G1, Canada}
\begin{abstract}


We propose a protocol by which entanglement can be extracted repeatedly from a quantum field. In analogy with prior work on entanglement harvesting, we call this protocol \emph{entanglement farming}. It consists of successively sending pairs of unentangled particles through an optical cavity. Using non-perturbative Gaussian methods, we show that in certain generic circumstances this protocol drives the cavity field towards a non-thermal metastable state. This state of the cavity is such that successive pairs of unentangled particles sent through the cavity will reliably emerge significantly entangled. We calculate thermodynamic aspects of the harvesting process, such as energies and entropies, and also the long-term behavior beyond the few-mode approximation. Significant for possible experimental realizations is the fact that this entangling fixed point state of the cavity is reached largely independently of the initial state in which the cavity was prepared. Our results suggest that reliable entanglement farming on the basis of such a fixed point state should be possible also in various other experimental settings, namely with the to-be-entangled particles replaced by arbitrary qudits and with the cavity replaced by a suitable reservoir system.

\end{abstract}

\maketitle

\section{Introduction}

Entanglement is a fundamental feature of quantum theory as well as a key resource for quantum computing and quantum communication. It is important, therefore, for fundamental research as well as for practical applications, to study mechanisms by which entanglement can be generated and transferred. Here, we will study the extent to which entanglement can be harvested sustainably from quantum fields by using localized quantum systems. The localized quantum systems can be, e.g., ions, atoms or molecules. They may also be idealized systems such as harmonic oscillators or 2-level Unruh-DeWitt detectors which have been shown, see e.g., \cite{Wavepackets}, to represent good models as long as no exchange of angular momentum is involved. For simplicity, we will here use the term \it detector \rm to refer to any localized quantum system used for entanglement harvesting from a field.    

Let us begin by recalling that a free quantum field may be regarded as a continuum of coupled harmonic oscillators filling space.
The fact that they are coupled implies that the ground state of these oscillators is an entangled state. In fact, the ground state contains quantum correlations at both non-spacelike and spacelike separations \cite{Unruh,summerswerner, summerswernerII}. 

Consider now two initially unentangled detectors in their ground states that start to interact with the vacuum. The two detectors generally become entangled. One mechanism by which they can become entangled is that they interact with each other causally, by exchanging information via the field. To this end the interactions of the detectors with the field must be non-spacelike so that the interaction between them can be mediated through the quantum field. The necessary energy for the existence of field quanta that mediate the detectors' interaction can come from two sources. Namely, the switching on of their interaction will make van der Waals-Casimir-Polder energy available \cite{CasPold,mathieuachim1}. Also, and generally more dominantly, the process of switching on and off the interaction breaks the time-translation invariance of the Hamiltonian, which via Noether's theorem can provide energy to the system by parametric forcing.  

Interestingly, the detectors can become entangled also if their interactions with the vacuum are spacelike separated, i.e., even if the two detectors have no opportunity to causally interact via the exchange of field quanta  \cite{Reznik1}. This is possible because the vacuum also possesses spacelike entanglement. A related phenomenon is that  timelike separated detectors can swap entanglement from the vacuum of a massless field \cite{Olson2011}, even though the field only propagates on the lightcone itself. Experimental realizations of these phenomena in superconducting circuits or quantum optical settings may soon become feasible \cite{Presabin,PastFutPRL}.

Here, we will study the harvesting of entanglement by two detectors that are initially unentangled and in their ground states and which then briefly travel through a cavity. The two detectors pick up entanglement by interacting with the cavity field. In particular, we will study the sustainability of this entanglement harvesting when passing pair after pair of fresh unentangled detectors through the cavity.  

Several competing phenomena occur in this case, raising several questions:  How efficient is it to utilize the swapping of entanglement from the vacuum at spacelike separations, compared to the non-spacelike separation case which utilizes detector interactions mediated via the cavity modes?  How does the entanglement extraction from the vacuum reported in \cite{Reznik1} compare with the way in which two quantum systems can be entangled via the interaction with a third system (see, e.g.,  Refs.~\cite{Paz2008,Zell2009})?  The periodic entering and exiting of the detectors will parametrically drive the cavity modes. In the long term (i.e., when continuing to send pairs of detectors through the cavity) will this excite and possibly heat up the cavity? Fresh detectors always arrive in their ground state, however. Does this lead to a cooling of the cavity?  What is the impact of entanglement monogamy on continued entanglement extraction? In the long term, will the cavity modes be driven towards or away from a thermal state? Will the cavity modes be driven towards a stable or at least a metastable state?  Finally, is the harvesting of entanglement by successive pairs of detectors pairs sustainable, i.e., do we have merely temporary harvesting of an in total small amount of entanglement or do we obtain actual farming of what will be large amounts of entanglement?   

To answer these questions, we will need to solve the dynamics of the cavity field in which we repeatedly inject detector pairs which are in their ground state. We let the detectors interact with the cavity field for a finite time, after which we remove the two detectors and insert in the cavity a new pair of detectors that are in their ground state. The state of the field is modified by each pair, which in turn changes how the field will affect new detectors placed in the cavity. 

To calculate the ability of successive pairs of detectors to harvest entanglement by passing through the cavity, we will use the Gaussian formalism applied to harmonic oscillator detectors \cite{Dragolegas,oscillator,Brusko}. In particular, we will employ the multimode non-perturbative techniques recently introduced in Ref.~\cite{oscillator}. We will choose particle detectors which act much like 2-level Unruh-DeWitt detectors but that are harmonic oscillators instead of 2-level systems. By doing so, the total system of the detectors and the field modes consists of exclusively coupled harmonic oscillators. This makes it possible to use  the Gaussian quantum mechanics techniques described in~\cite{oscillator} to solve for the evolution of the detector-field system non-perturbatively in the field-detector coupling. 

In this way we go beyond the usual perturbative approach based on the two-level Unruh-DeWitt detector model which is widely used in the literature, among others in the original derivation of entanglement harvesting from the vacuum \cite{extraction}. 
By doing so, we find that the dynamics of the detectors plus field system can be expressed as a linear dynamical system, allowing us to study the system's dynamics for arbitrarily long times, i.e., for a large number of detector pairs passing through the cavity.

Our results will show that keeping the interaction of the detectors in the cavity short enough to be spacelike separated is not efficient: it is not sustainable to extract entanglement over many cycles and the total amount that can be extracted is small. 

As one might expect, allowing the interaction of the detectors in the cavity to be long enough to become non-spacelike (i.e., allowing the detectors to interact via the cavity field) makes the detectors' entanglement harvesting more efficient. 

What is surprising is the magnitude of the amount of entanglement that can be harvested by repeated detector insertions. Namely,  we will first find that over repeated cycles of detector pair insertions, a cavity field that is initially in the vacuum state is driven towards a metastable, highly non-thermal state. Even a cavity field that is initially in a thermal state, from which no entanglement can be harvested, is driven towards that metastable state.
  
Second, we will find that this metastable state allows each new pair of detectors to become significantly entangled. Therefore,  by repeating this protocol, one would obtain a stream of significantly but not maximally entangled detectors, whose entanglement could then, for example, be distilled into maximally entangled EPR pairs. The metastable state is very long-lived under repeated entanglement harvesting by detector pairs, although this state eventually does turn into a non-entangling state. This happens when higher, off-resonant modes start to become significantly excited, and hence it is beyond the single-mode approximation. In practice, if the excitations of those higher modes slightly leak from the cavity, fully sustainable entanglement farming should be possible. 

\section{Gaussian quantum mechanics}

As we mentioned above, we will use the formalism of Gaussian quantum mechanics and will therefore model the detectors as quantum harmonic oscillators \cite{oscillator}. This approach is particularly well suited for the scenario considered here because the entire evolution, including the step of replacing the used detectors with a fresh pair each cycle, can be naturally and simply expressed in terms of a covariance matrix.
Furthermore, with this model we will be able to solve for the system evolution completely non-perturbatively and analytically, namely by expressing it as a matrix exponential.

We begin with a brief overview of Gaussian quantum mechanics, focusing on the concepts necessary for the subsequent sections. A more complete introduction to the Gaussian formalism is given by \cite{adesso}, among many other resources available in the literature. Unless otherwise stated full derivations of the content in subsections \ref{GaussBasics} and \ref{GaussEnt} can be found in \cite{adesso}. In subsection \ref{sectRel} we also discuss the relative entropy between Gaussian states including a full derivation.

\subsection{Basics}   \label{GaussBasics}

Consider a set of continuous-variable bosonic modes. These can describe, for example, a set of $N$ quantum harmonics oscillators or the modes of a field (or both, as we consider here). We label the quadrature operators for each mode $(\hat{q}_i,\hat{p}_i)$ for each $i=1 \dots N$, where these operators satisfy the canonical commutation relations $[\hat{q}_i,\hat{p}_j]=i\delta_{ij}$. In later sections, we will distinguish between the two oscillators belonging to the detectors and the oscillators belonging to the modes of the field.  In this section, however, it will be convenient not to distinguish them and to package all these degrees of freedom into a phase-space vector of the form:
\begin{align}
	\opvec{x}=(\hat{q}_1,\hat{p}_1, \hat{q}_2, \hat{p}_2, \dots, \hat{q}_N,\hat{p}_N)^T.
\end{align}
These quadrature operators are related to the creation and annihilation operators of each mode by
\begin{align}
	\hat{q}_i=\frac{1}{\sqrt{2}}(\hat{a}_i+\hat{a}^\dagger_i), \;\;\;\; \hat{p}_i=\frac{i}{\sqrt{2}}(\hat{a}_i^\dagger-\hat{a}_i).
\end{align}

A Gaussian state of such a system is one that can be completely described by its first and second moments. Equivalently, its Wigner function takes the form of a Gaussian. We will here use Gaussian states that have zero mean. These can be fully described by their covariance matrix $\mat{\sigma}$, the entries of which are\footnote{Note that this definition differs by a factor of two from what is usually called the covariance matrix. Our definition offers the advantage that the vacuum state is represented by the identity matrix.}
\begin{align} \label{covmat}
	\sigma_{i j}=\langle \hat{x}_i \hat{x}_j+\hat{x}_j \hat{x}_i \rangle.
\end{align}
The formalism of Gaussian states can be extended to states with a non-zero mean, but we will not need that generalization.
Gaussian states account for a wide range of states that are of interest in quantum information, quantum optics, and relativistic field theory; they include vacuum, thermal, squeezed, and coherent states.

Specifying the reduced state of a subset of modes is particularly straightforward when working with Gaussian states. One must simply use the rows and columns of the total covariance matrix that corresponds to the modes one wishes to isolate. For example the reduced state of the first two modes of our ensemble, represented by the covariance matrix $\mat{\sigma}_{12}$, is simply given by the first $4 \times 4$ block of $\mat{\sigma}$. Furthermore, this reduced state will be of the form
\begin{align} \label{twomode}
	\mat{\sigma}_{12}=
	\begin{pmatrix}
		\mat{\sigma}_1 & \mat{\gamma}_{1 2} \\
		\mat{\gamma}_{1 2}^T & \mat{\sigma}_2
	\end{pmatrix},
\end{align}
where the $2 \times 2$ covariance matrices $\mat{\sigma}_1$ and $\mat{\sigma}_2$ represent the reduced states of the first and second modes on their own, respectively. The $2\times 2$ matrix $\mat{\gamma}_{12}$ encodes the correlations between the two modes; they are in a product state with respect to each other iff $\mat{\gamma}_{12}=0$.

In our study we use the fact that any unitary evolution generated by a quadratic Hamiltonian will preserve the Gaussianity of states \cite{schumaker}. Any such unitary operation $\hat{U}$ on the Hilbert space of states corresponds to a symplectic transformation on the phase space that, in the Heisenberg picture, transforms the quadrature operators via $\opvec{x} \rightarrow \hat{U}^\dagger \opvec{x} \hat{U}=\mat{S}\opvec{x}$. Here $\mat{S}$ is a symplectic matrix, which satisfies the condition
\begin{align} \label{symplectic}
	\mat{S} \mat{\Omega} \mat{S}^T=\mat{S}^T \mat{\Omega} \mat{S}=\mat{\Omega},
\end{align}
where $\mat{\Omega}$ is the symplectic form, defined as
\begin{align} \label{symform}
	\mat{\Omega}=\bigoplus_{i=1}^N
	\begin{pmatrix}
		0 & 1 \\
		-1 & 0
	\end{pmatrix}.
\end{align}
The condition \eqref{symplectic} is equivalent to the requirement that the canonical commutation relations $[\hat{q}_i,\hat{p}_j]=i\delta_{ij}$ are preserved throughout the evolution. Acting on the covariance matrix this evolution is given by $\mat{\sigma} \rightarrow \mat{S} \mat{\sigma} \mat{S}^T$. Thus if we have some unitary evolution in time this can be represented by a time-dependent symplectic matrix $\mat{S}(t)$, and the state of our system as a function of time is fully determined by the covariance matrix
\begin{align}  \label{covmatevol}
	\mat{\sigma}(t)=\mat{S}(t)\mat{\sigma}_0 \mat{S}(t)^T,
\end{align}
where $\mat{\sigma}_0$ represents the system's initial state.

The free Hamiltonian associated with our system of oscillators/modes is given by
\begin{align} \label{freeH}
	\op{H}_\text{free}=\sum_{i=1}^N \omega_i \hat{a}_i^\dagger \hat{a}_i=\sum_{i=1}^N \frac{\omega_i}{2}(\hat{p}_i^2+\hat{q}_i^2),
\end{align}
where we have ignored any constant addition to $\hat{H}_\text{free}$ since this will have no impact on the states or evolution considered. From this, we immediately see that the expected energy of any free Gaussian state can be computed via
\begin{align} \label{energy}
	E=\langle \hat{H}_\text{free} \rangle=\sum_{i=1}^N \frac{\omega_i}{2} \text{Tr} \mat{\sigma}_i,
\end{align}
where $\mat{\sigma}_i$ are the covariance matrices of the individual modes.

The ground/vacuum state of the Hamiltonian Eq. (\ref{freeH}) is Gaussian, and can be shown to have a covariance matrix given simply by the $2N \times 2N$ identity matrix:
\begin{align}
	\mat{\sigma}_\text{vac}=\mat{I}_{2N}.
\end{align}
Because the free Hamiltonian is devoid of coupling between modes, this vacuum state is equivalent to a product of single modes, each of which are in their ground state.

We will also be interested in considering thermal Gibbs states of this Hamiltonian. Such a state is also Gaussian, and at temperature $T$ has a covariance matrix of the form
\begin{align} \label{thermalstate}
	\mat{\sigma}_\text{therm}= \bigoplus_{i=1}^N
	\begin{pmatrix}
		\nu_i^{(T)} & 0 \\
		0 & \nu_i^{(T)}
	\end{pmatrix},
\end{align}
where
\begin{align}
	\nu_i^{(T)}=\frac{\exp{\omega_i \beta}+1}{\exp{\omega_i \beta}-1}, \;\;\;\; \beta \equiv 1/T.
\end{align}
Similar to the vacuum, a thermal state is a product state of individual modes, each of which is are in a thermal state at temperature $T$.

The values $\nu_i^{(T)}$ are known as the symplectic eigenvalues of a thermal state. More generally the set of $N$ symplectic eigenvalues $\nu_i$ of a given Gaussian state $\mat{\sigma}$ consists of the eigenvalues of the matrix $|i\mat{\Omega}\mat{\sigma}|$.
The symplectic eigenvalues characterize Gaussian states up to symplectic transformations, and any Gaussian state can be symplectically transformed to a form with its symplectic eigenvalues placed in pairs along its diagonal, with zeros in any off-diagonal entries. This is called the Williamson normal form. A thermal state  Eq.~\eqref{thermalstate}, for example, is already in its Williamson normal form.

The symplectic eigenvalues satisfy $\nu_i \geq 1$ by the uncertainty principle, and a Gaussian state is pure if any only if $\nu_i=1, \; \forall \; i$ (meaning that a pure Gaussian state necessarily saturates the uncertainty principle).
Conversely, a mixed Gaussian state will have one or more of these values greater than unity. 
A quantitative measure of the degree of purity of the state is given by the purity
$P=\prod_i \nu_i^{-1}$. 
Since the number $\det \mat{\sigma}$ is a symplectically invariant quantity, it follows that the purity of a Gaussian state $\mat{\sigma}$ is simply
\begin{align} \label{purity}
	P=\frac{1}{\sqrt{\det \mat{\sigma}}}.
\end{align}

The symplectic eigenvalues of $\mat{\sigma}$ also determine the von Neumann entropy of the state. 
From the $N$ symplectic eigenvalues $\nu_i$ the von Neumann entropy is given by
\begin{align}   \label{entropy}
	S(\mat{\sigma})=\sum_{i=1}^N f(\nu_i),
\end{align}
where
\begin{align}
	f(x) \equiv \frac{x+1}{2}\log \left(\frac{x+1}{2} \right)-\frac{x-1}{2}\log \left(\frac{x-1}{2} \right).
\end{align}

\subsection{Entanglement}   \label{GaussEnt}

We will need to know how to compute the entanglement in a two-mode state $\mat{\sigma}_{12}$, which will be of the form Eq. (\ref{twomode}). 
In the case that $\mat{\sigma}_{12}$ is pure, the natural measure of entanglement is the entanglement entropy $S(\mat{\sigma}_1)=S(\mat{\sigma}_2)$. 
However, generically the state of the detectors is mixed, and we must use a different measure of entanglement.

Here we will adopt the logarithmic negativity $E_N$ as an entanglement measure \cite{plenio}.
The logarithmic negativity will be particularly appropriate for our study since it has the property of being additive on product states: $E_N (\rho \otimes \tilde{\rho})=E_N(\rho)+E_N(\tilde{\rho})$, or when using covariance matrices: $E_N (\mat{\sigma} \oplus \tilde{\mat{\sigma}})=E_N(\mat{\sigma})+E_N(\tilde{\mat{\sigma}})$. For two-mode states the presence of logarithmic negativity is both necessary and sufficient for the presence of entanglement. Here we will only quote its form in the Gaussian formalism, but the full derivation can be found in \cite{adesso}. For a state $\mat{\sigma}_{1 2}$ of the form Eq. (\ref{twomode}) the logarithmic negativity is given by
\begin{align} \label{logneg}
	E_N=\max (0, -\log \tilde{\nu}_i),
\end{align}
where $\tilde{\nu}_i$ is the smaller of the state's partially transposed symplectic eigenvalues, which can be computed from
\begin{align}
	2\tilde{\nu}_-^2=\tilde{\Delta}-\sqrt{\tilde{\Delta}^2-4 \det \mat{\sigma}_{1 2}},
\end{align}
where $\tilde{\Delta}=\det \mat{\sigma}_1+\det \mat{\sigma}_2-2\det \mat{\gamma}_{1 2}$. Notice that these are \emph{not} generally the same as the usual symplectic eigenvalues.

\subsection{Relative entropy}  \label{sectRel}

Lastly, we will find it useful to consider the relative entropy between two Gaussian states. 
As the derivation of this is not common in the literature we will include a full derivation here. 
Please note that, to avoid notational confusion, we maintain that covariance matrices will be boldfaced whereas density matrices will not be.

The relative entropy of two density matrices $\rho_A$ and $\rho_B$ is defined as
\begin{align}
S(\rho_A || \rho_B) &= \text{Tr} \rho_A (\log \rho_A - \log \rho_B).
\end{align}
The relative entropy provides a measure of distinguishability of quantum states \cite{Vedral2001}.

An important property of the relative entropy is \emph{monotonicity}.
Under any trace-preserving completely positive map $\Phi$ we have
\begin{equation}
S(\Phi(\rho_A) || \Phi(\rho_B)) \leq S(\rho_A || \rho_B).
\end{equation}
In particular, if we choose $\rho_B$ to be a fixed point of the evolution ($\Phi(\rho_B) = \rho_B$) then we have
\begin{equation}
S(\Phi(\rho_A) || \rho_B) \leq S(\rho_A || \rho_B)
\end{equation}
so that we have a measure of the distance between $\rho_A$ and the equilibrium state that is monotonic along the evolution.
It therefore provides a measure of the degree to which the system has equilibrated.

Let us consider the case where the states $\rho_A$ and $\rho_B$ are Gaussian, with covariance matrices $\mat{\sigma}^A$ and $\mat{\sigma}^B$ respectively:
\begin{equation}
\sigma^A_{ij} = \left \langle x_i x_j + x_j x_i \right \rangle_{\rho_A}, \qquad 
\sigma^B_{ij} = \left \langle x_i x_j + x_j x_i \right \rangle_{\rho_B}.
\end{equation}
To compute the relative entropy, we express it as 
\begin{equation}
S(\rho_A || \rho_B) = -S(\rho_A) - \langle \log \rho_B \rangle_{\rho_A}.
\end{equation}
The entropy term is easily computed from the symplectic eigenvalues of $\mat{\sigma}^A$, as given by Eq. (\ref{entropy}). Since $\rho_B$ is a Gaussian state, $\log \rho_B$ is a quadratic operator, and its expectation value is easily expressed in terms of the covariance matrix.

To find $\log \rho_B$, we first symplectically diagonalize $\mat{\sigma}^B$. That is, we put it into its Williamson normal form.
Let us choose modes $\tilde x_i = \sum_j S_{ij} \op{x}_j$ such that
\begin{equation}
\langle \tilde x_i  \tilde x_j  + \tilde x_j  \tilde x_i \rangle_{\rho_B} =\sum_{kl} S_{ik} S_{jl} \sigma^B_{kl} = D_{ij}
\end{equation}
where $\mat{D} = \text{diag}(\nu_1, \nu_1, \nu_2, \nu_2, \ldots)$, with symplectic eigenvalues $\nu_1, \ldots, \nu_N$.
In this basis the density matrix $\rho_B$ takes the form \cite{adesso}
\begin{equation} \label{sigma}
\rho_B = \bigotimes_{k = 1}^N \frac{2}{\nu_k + 1} \sum_{n \geq 0} \left ( \frac{\nu_k-1}{\nu_k+1} \right)^n \ket{n}_k \prescript{}{k}{\bra{n}}.
\end{equation}
Here $|n \rangle_k$ are the eigenstates of the number operator 
\begin{equation}
N_k = \frac{1}{2} \left( \tilde x_{2k}^2 + \tilde x_{2k+1}^2 \right) - \frac{1}{2}.
\end{equation}
Taking the logarithm of Eq.~\eqref{sigma}, we find that
\begin{align}
\log \rho_B &= \sum_{k=1}^N \left( \log \frac{2}{\nu_k + 1} + \log \frac{\nu_k - 1}{\nu_k + 1} N_k \right) \\
&= c + \sum_{i,j=1}^{2N} \tilde H_{ij} \tilde x_i \tilde x_j
\end{align}
where we have defined
\begin{align}
c &= \sum_{k} \frac{1}{2} \log \frac{4}{\nu_k^2 - 1}, \label{c} \\
\tilde H &= \text{diag}(H_1, H_1, H_2, H_2, \ldots), \\
H_k &= \frac{1}{2} \log \frac{\nu_k - 1}{\nu_k + 1}.
\end{align}
Changing basis from the $\tilde x$ basis back to the $\op{x}$ basis $\log \rho_B$ takes the form
\begin{equation}
\log \rho_B = c + \sum_{i,j=1}^{2N} H_{ij} x_i x_j, \qquad H_{kl} = S_{ik} S_{jl} \tilde H_{ij}. \label{H}
\end{equation}
The relative entropy is therefore given by
\begin{equation}
S(\rho_A || \rho_B) = - S(\rho_A) - c - \frac{1}{2} \sum_{i,j=1}^{2N} \sigma^A_{ij} H_{ij},
\end{equation}
with $c$ and $H$ given by Eqs. \eqref{c} and \eqref{H}.
The quantity $S(\rho_A)$ can be found by Eq. (\ref{entropy}) in terms of the symplectic eigenvalues of the matrix $\mat{\sigma}^A$.

When the state $\rho_B$ is a Gibbs thermal state 
\begin{equation}
\rho_B = Z^{-1} e^{- \beta \op{H}},
\end{equation} 
of inverse temperature $\beta$ with respect to some given Hamiltonian $\op{H}$, the relative entropy reduces to the free energy difference 
\begin{equation} \label{thermal_relative_entropy}
S(\rho_A || \rho_B) = \beta ( F(\rho_A) - F(\rho_B) )
\end{equation}
where the free energy is defined as
\begin{equation}
F(\rho_A) = \langle \op{H} \rangle_{\rho_A} - \beta^{-1} S(\rho_A).
\end{equation}

We will be interested in knowing, for a given Gaussian state $\rho_A$, what thermal state $\rho_B$ is closest to $\rho_A$. For this task we will use the relative entropy as our distance measure.
To find the thermal state that minimizes the relative entropy \eqref{thermal_relative_entropy}, we simply extremize with respect to the inverse temperature $\beta$:
\begin{equation}
\frac{d}{d \beta} S(\rho_A || \rho_B) = \langle \op{H} \rangle_{\rho_A} - \langle \op{H} \rangle_{\rho_B} = 0.
\end{equation}
Thus the thermal state $\rho_B$ closest to the given state is that which has the same energy. Recall that the energy of a Gaussian state is easily computed via Eq. (\ref{energy}).
When the temperature of the state $\rho_B$ is chosen in this way, the relative entropy is straightforwardly shown to be simply the entropy difference 
\begin{equation}  \label{thermalrel}
S(\rho_A || \rho_B) = S(\rho_B) - S(\rho_A).
\end{equation}
The fact that the relative entropy is non-negative reflects the fact that the thermal state is the unique state of maximal entropy among states of a given energy.

Thus we establish an ``effective temperature'' $\beta^{-1}$ of our state, obtained by solving for $\beta$ as a function of the energy \footnote{While we cannot find $\beta$ analytically, it can be easily found numerically because the energy is monotonically increasing with temperature.}, and a measure of distance to the manifold of thermal states.

\section{The model and evolution}   \label{modelSect}

We consider the simple scenario of a massless scalar field in a one-dimensional reflecting cavity of length $L$. The field at the boundaries are constrained to be zero, thus giving us Dirichlet boundary conditions. In practice we will need to apply a UV cutoff to the field, namely to only include a finite number of field modes. Doing so is perfectly acceptable as long as the detector-field interaction strength is not too strong. We have been vigilant in our study to include enough field modes in our system, such that adding additional modes does not modify our results. We will let the number of field modes included be $N-2$, such that with two detectors the total number of degrees of freedom is $N$. We will find it most convenient to work in the Schr\"odinger picture, in which case our cavity field takes the simple form \cite{BandD}
\begin{align}
	\hat{\phi}(x)=\sum_{n=1}^{N-2}\frac{1}{\sqrt{\pi n}}(\hat{a}_n+\hat{a}_n^\dagger)\sin (k_n x), \;\;\;\; 0<x<L,
\end{align}
where $k_n=n \pi/L$. Here $\hat{a}_n^\dagger$ and $\hat{a}_n$ are the creation and annihilation operators of the field modes.

We now ask what happens when a pair of harmonic oscillator detectors in their ground state and each with characteristic frequency $\Omega$ are injected into the cavity at positions $x_1$ and $x_2$, made to interact with the field for a time of $t_f$, then removed (where they can be examined and/or used) and replaced with a fresh new pair of detectors that are in their ground state, and then have this process repeated many times. We will first describe how to solve for the evolution of the detectors+field system, and then will go on to discuss how to implement the cycles of injecting and removing detectors. Both of these procedures are straightforward given the machinery of the oscillator detectors and Gaussian quantum mechanics \cite{oscillator}.

In order to be able to use the Gaussian formalism we have to choose an interaction between the detectors and the field whose Hamiltonian is no more than quadratic in quadrature operators, as such Hamiltonians generate Gaussianity-preserving evolution \cite{schumaker}. We will here choose the monopole-monopole type coupling that has been widely adopted in Unruh-DeWitt detector models \cite{dewitt}. The corresponding interaction Hamiltonian (again, in the Schr\"odinger picture) between the field and two stationary detectors at positions $x_1$ and $x_2$ is
\begin{align} \label{Hint}
	\hat{H}_\text{int}= \lambda_1 (\hat{a}_{d1}+\hat{a}_{d1}^\dagger)\hat{\phi}(x_1)+ \lambda_2 (\hat{a}_{d2}+\hat{a}_{d2}^\dagger)\hat{\phi}(x_2),
\end{align}
where $\hat{a}_{d1}$ and $\hat{a}_{d2}$ are the annihilation operators of the detectors and the numbers $\lambda_1$ and $\lambda_2$ are the coupling strengths; in this paper we will consider them to be equal for both detectors, $\lambda_1=\lambda_2=\lambda$. Note that the case in which the detectors are not stationary is easily represented; one simply replaces $x_{1,2}$ with the detector positions as a function of time \cite{oscillator}.

Since we are working in the Schr\"odinger picture the state evolution will be generated by the total Hamiltonian, including both the free and interacting parts: $\hat{H}=\hat{H}_\text{free}+\hat{H}_\text{int}$. Here the free Hamiltonian includes contributions from both the detectors and the field,
\begin{align} \label{Hfree2}
	\hat{H}_\text{free}=\Omega \hat{a}_{d1}^\dagger \hat{a}_{d1}+\Omega \hat{a}^\dagger_{d2} \hat{a}_{d2}+\sum_{n=1}^{N-2} \omega_n \hat{a}_n^\dagger \hat{a}_n,
\end{align}
where $\Omega$ is the frequency of the detectors and $\omega_n=k_n$ since the field is massless.

To solve for the evolution generated by $\hat{H}=\hat{H}_\text{free}+\hat{H}_\text{int}$, let us order the phase space of the detectors and field modes into a vector of the form
\begin{align}
	\opvec{x}=(\hat{q}_{d1},\hat{p}_{d1}, \hat{q}_{d2}, \hat{p}_{d2}, \hat{q}_1,\hat{p}_1, \dots, \hat{q}_{N-2},\hat{p}_{N-2})^T,
\end{align}
where the first $4$ entries correspond to the detectors, and the rest to the field. The covariance matrix $\mat{\sigma}$ of the detectors+field system then follows as normally via Eq. (\ref{covmat}). The Hamiltonian $\hat{H}$ will generate some time-dependent symplectic evolution given by a symplectic matrix $\mat{S}(t)$ such that from an initial state $\mat{\sigma_0}$ the time-evolved state is represented by $\mat{\sigma}(t)=\mat{S}(t)\mat{\sigma}_0 \mat{S}(t)^T$, Eq. (\ref{covmatevol}).

Any (in general time-dependent) quadratic Hamiltonian can then be represented as a Hermitian matrix $\mat{F}(t)$ such that
\begin{align}
	\hat{H}(t)=\opvec{x}^T \mat{F}(t) \opvec{x}.
\end{align}
From this, the symplectic evolution $\mat{S}(t)$ generated by $\hat{H}(t)$ can be shown to solve the matrix equation \cite{oscillator}
\begin{align} \label{EOM}
	\frac{d}{dt} \mat{S}(t)=\mat{\Omega} \mat{F}^\text{sym}(t) \mat{S}(t),
\end{align}
where $\mat{F}^\text{sym}=\mat{F}+\mat{F}^T$ and $\mat{\Omega}$ is the symplectic form, Eq. (\ref{symform}). In the case where the detectors are stationary inside the cavity the Hamiltonian will be time-independent, as in Eq. (\ref{Hint}). This is the simplified scenario that we will be considering here, in which case the solution to Eq. (\ref{EOM}) is of the form
\begin{align} \label{soln}
	\mat{S}(t)=\exp (\mat{\Omega} \mat{F}^\text{sym} t).
\end{align}
In time-dependent scenarios Eq. (\ref{EOM}) can generally only be solved numerically.

Working in the Schr\"odinger picture we must include both the free and interaction Hamiltonians in the evolution equation. The symmetrized Hamiltonian matrix can thus be decomposed into the form $\mat{F}^\text{sym}=\mat{F}^\text{sym}_\text{free}+\mat{F}^\text{sym}_\text{int}$. For our particular scenario, as given by Eqs. (\ref{Hint},\ref{Hfree2}), we have explicitly $\mat{F}^\text{sym}_\text{free}=\text{diag}(\Omega,\Omega,\Omega,\Omega, \omega_1, \omega_1, \dots, \omega_{N-2},\omega_{N-2 })$ and
\begin{align}
	\mat{F}^\text{sym}_\text{int}=2\lambda
	\begin{pmatrix}
		\mat{0}_{4} & \mat{X} \\
		\mat{X}^T & \mat{0}_{2(N-2)}
	\end{pmatrix},
\end{align}
where $\mat{0}_n$ is the $n \times n$ matrix of zeros, and
\begin{align}
	\mat{X} \equiv
	\begin{pmatrix}
		\frac{\sin k_1 x_1}{\sqrt{\pi}} & 0 & \frac{\sin k_2 x_1}{\sqrt{2 \pi}} & 0 & \dots & 0 & \frac{\sin k_{N-2} x_1}{\sqrt{(N-2)\pi}} \\
		0 & 0 & 0 & 0 & \dots & 0 & 0 \\
		\frac{\sin k_1 x_2}{\sqrt{\pi}} & 0 & \frac{\sin k_2 x_2}{\sqrt{2 \pi}} & 0 & \dots & 0 & \frac{\sin k_{N-2} x_2}{\sqrt{(N-2)\pi}} \\
		0 & 0 & 0 & 0 & \dots & 0 & 0
	\end{pmatrix}.
\end{align}

Once the symplectic evolution matrix $\mat{S}(t)$ has been obtained via Eq. (\ref{soln}) the process of cycling pairs of detectors in and out of the cavity is trivially performed. Note critically that $\mat{S}(t)$ is independent of the initial state of the system. We thus need only compute the evolution once and can then use the same matrix for every cycle. The process is as follows. We begin with a cavity field in a state of our choosing, $\mat{\sigma}_f^{(0)}$ and a pair of detectors in a state of our choosing, $\mat{\sigma}_d^{(0)}$. Assuming that the field and detectors are initially uncorrelated (this need not be assumed if one were interested in such a scenario) then the total system state is initially $\mat{\sigma}^{(0)}\equiv \mat{\sigma}_d^{(0)} \oplus \mat{\sigma}_f^{(0)}$. We then inject the detectors into the cavity at time-zero; this sharply couples the detectors with the field by turning on the interaction Hamiltonian. The system is then left to evolve for a given time $t_f$. This evolution is governed by $\mat{S}(t_f)$, and so after this time the system is in a state of the form
\begin{align}
	\mat{\sigma}^{(1)} \equiv \mat{S}(t_f) \mat{\sigma}^{(0)} \mat{S}(t_f)^T \equiv
	\begin{pmatrix}
		\mat{\sigma}_d^{(1)} & \mat{\gamma}_{df}^{(1)} \\
		\mat{\gamma}_{df}^{(1)T} & \mat{\sigma}_f^{(1)}
	\end{pmatrix}.
\end{align}
Here $\mat{\sigma}_d^{(1)}$ and $\mat{\sigma}_f^{(1)}$ are the reduced states of the detectors and field, respectively. The matrix $ \mat{\gamma}_{df}^{(1)}$ encodes the correlations generated between the detectors and field during the evolution. The state of the detectors can now be examined and, for example, be tested for entanglement via Eq. (\ref{logneg}).

After this first stage of evolution we immediately remove the detectors from the cavity. They can then be sent elsewhere to be utilized, and for the current protocol we ignore any correlations between the pair and the field. As soon as the first pair is removed we inject into the cavity a fresh new pair of detectors which have been prepared in the same way as the first pair (although this need not be the case). The state of the new detectors+field thus becomes $\mat{\sigma}_d^{(0)}\oplus \mat{\sigma}_f^{(1)}$. We then allow this system to evolve for time $t_f$ (we could of course choose a different time if we wanted), after which we have the state $\mat{\sigma}^{(2)}=\mat{S}(t_f) [\mat{\sigma}_d^{(0)}\oplus \mat{\sigma}_f^{(1)}] \mat{S}(t_f)^T$. As before, the detectors-block of this matrix can be examined and/or utilized. We then remove these detectors and immediately replace them with a new pair, etc. After $k$ iterations of this process the detectors-field system can be defined recursively as
\begin{align}
	\mat{\sigma}^{(k)}=\mat{S}(t_f) [\mat{\sigma}_d^{(0)}\oplus \mat{\sigma}_f^{(k-1)}] \mat{S}(t_f)^T.
\end{align}
Each cycle we lose the information about the detectors-field correlation, and we consider each produced pair of detectors to be their own entity and resource. If desired, it is straightforward to store this correlation information in the $\mat{\gamma}_{df}^{(k)}$'s, to possibly utilize it in other protocols.

\section{Results}

Here we present our primary findings. 
First, in subsection \ref{fixed} we characterize the evolution of the field in terms of a Gaussian superoperator: a linear map on the space of covariance matrices of the cavity field.
As a first step, we will initially ignore the contribution of highly off-resonant modes, i.e., we only include modes with frequencies within some resonant window centered on $\Omega$. Up to a very large number of cycles this produces a very good approximation for the evolution of the injected detectors.
We find that over successive cycles, the cavity field rapidly converges toward a fixed point which is independent of the initial state of the field. Asymptotically approaching the fixed point, significant detector-detector entanglement is  obtained every cycle. The cavity then effectively acts as a stable medium through which pairs of detectors can acquire entanglement.

However, when also the highly off-resonant modes are considered, we find that this fixed point becomes  unstable after a very large number of cycles. Namely, after many iterations of the protocol, the off-resonant field modes diverge towards increasingly energetic and mixed states. Even though these modes are effectively very weakly coupled to the detectors because they are highly off-resonant, they eventually become sufficiently excited to have enough of a decohering effect that the generation of entanglement between the detectors is impeded. While the detectors will be interacting with the near resonant modes in the same way as they have previously, they will also now experience a highly energetic and highly populated photon gas. The detectors' interactions with this high-mode photon gas evidently introduces enough decoherence into their evolution that any entanglement garnered per cycle will be lost. 

Interestingly, the timescale associated to this instability is long enough to allow an extremely large quantity of entanglement to be extracted. For example, depending on the chosen parameters this could easily be on the order of $10^5$ total amount of logarithmic negativity, when added up over all detector pairs.
Notice that the instability is an effect which is beyond the single-mode approximation that is commonly used in quantum optics, and it occurs independently of how long each cycle is. We will discuss this effect in more detail in Sec. \ref{Sectinstability}.

Unless otherwise stated, all data presented in this section was obtained using the following parameters: $\lambda=0.01$, $L=8$, $\Omega=\pi/8$ (resonant with the fundamental mode), $x_1=L/3$, and $x_2=2L/3$ (such that the distance between the detectors is $r=8/3$). There is nothing special about this choice and similar (correspondingly scaled) results would be obtained for different values of the parameters.

\subsection{Fixed point analysis}   \label{fixed}

A key question that we aim to answer in this paper is how many times a cavity may be reused for entanglement extraction.
We therefore now consider how the field evolves under repeated application of the entanglement extraction protocol.
The field evolution is expressed as a Gaussian superoperator, i.e. a linear map acting on the field covariance matrix.
This allows us to treat the field evolution as a linear discrete dynamical system, and to characterize the evolution.
In particular we will find regimes in which there is an evolution toward a long-lived metastable state from which entanglement can be extracted.

As described above, one interaction cycle consists of three steps: introducing two new detectors in their ground states, evolving for a fixed amount of time under a quadratic Hamiltonian, and tracing over the detector Hilbert space.
Let $\mat{\sigma}_f^{(k)}$ denote the state of the field after $k$ iterations of the cycle have already occurred, and let $\mat{S}$ be the symplectic transformation describing the system evolution over a single cycle, as given by Eq. (\ref{soln}). We can express this transformation in block form as
\begin{equation}
\mat{S} = \left[ \begin{array}{cc} \mat{A} & \mat{B} \\ \mat{C} & \mat{D} \end{array} \right],
\end{equation}
where $\mat{A}$ is a $4$-dimensional matrix and $\mat{D}$ is $2(N-2)$-dimensional. After an interaction cycle, the field will be left in the state
\begin{equation} \label{field_evolution}
\mat{\sigma}_f^{(k+1)} = \mat{D} \mat{\sigma}_f^{(k)} \mat{D}^T + \mat{C} \mat{C}^T.
\end{equation}
We call the map described by Eq.~\eqref{field_evolution} the Gaussian superoperator.

We can cast this into a more familiar form if we view the matrix $\mat{\sigma}_f^{(k)}$ as a vector $\vec v^{(k)}$ in the symmetric subspace of the tensor product $\mathbb{R}^{2 (N-2)} \otimes \mathbb{R}^{2(N-2)}$ where $N-2$ is the number of field modes. 
Equation \eqref{field_evolution} then takes the form
\begin{equation} \label{dynamical_system}
\vec v^{(k+1)} = (\mat{D} \otimes \mat{D}) \vec v^{(k)} + \vec{c}
\end{equation}
where $\vec{c}$ is the vector corresponding to the matrix $\mat{C} \mat{C}^T$. 

Eq.~\eqref{dynamical_system} is an affine discrete dynamical system, and its dynamics is characterized by the eigenvalues of the symmetric tensor product $\mat{D} \otimes_S \mat{D}$.
These are completely determined by the eigenvalues of $\mat{D}$. 
Let $d_1, \ldots, d_{2N}$ be the eigenvalues of $\mat{D}$, with $\abs{d_1} \geq \abs{d_2} \geq \cdots$. 
The eigenvalues of $\mat{D} \otimes_S \mat{D}$ are given by $\{ d_i d_j : i \leq j \}$.
Thus we can determine aspects of the dynamics of the dynamical system \eqref{dynamical_system} from the eigenvalues of the matrix $\mat{D}$.

If the eigenvalues of $\mat{D}$ (and therefore those of $\mat{D} \otimes_S \mat{D}$) are all within the unit circle, then the field is driven toward a unique fixed point, given by
\begin{align}  \label{fixedsol}
\vec v_\text{fixed}=(\mat I - \mat{D} \otimes \mat{D})^{-1}\vec c.
\end{align}
By converting this vector back into a matrix we then obtain the fixed point covariance matrix $\mat{\sigma}_\text{fixed}$.

The rate at which such a fixed point is approached is determined by the largest modulus eigenvalue $d_1$ of $\mat{D}$.
The eigenvector corresponding to eigenvalue $d_1$ will converge exponentially to the fixed point in a number of cycles on the order of
\begin{equation}
n = - 1 / \log | d_1 |.
\end{equation}
A generic initial state will have a nonzero projection in this direction, and therefore its approach to the fixed point is determined by eigenvalue with the largest modulus.

In the case where $\mat{I} -\mat{D} \otimes \mat{D}$ is not invertible, there is a subspace of fixed points.
This occurs when there are modes that are completely uncoupled with the detectors, such as in the case where both detectors sit at nodes of the corresponding mode functions.
There can also be approximate fixed points for modes that are weakly coupled to the detectors, as in the case of modes that are far from resonance.
Since there are many off-resonant modes, our system has a large space of approximate fixed points.

If $\abs{d_1} > 1$ the system has no absolutely stable fixed point. 
The initial covariance matrix can be expanded in the eigenbasis of the superoperator, and generically the projection onto the eigenspace with eigenvalue $d_1$ will be nonzero.
If this is the case, the covariance matrix will grow exponentially in some direction in the space of symmetric matrices, and the time scale of this growth (in number of iterations) is
\begin{equation} \label{instability}
n = 1 / \log \abs{d_1}.
\end{equation}
Although the growth of the covariance matrix depends on the initial state, this dependence is linear while the dependence on $d_1$ is exponential in the number of iterations.
Thus in practice we find that Eq.~\eqref{instability} yields a good indication of the timescale for instability.

This instability has a strong effect on the process of entanglement extraction.
After interaction with the field, the covariance matrix of the detectors is linear in the initial state of the field.
Even though the growing modes of the field interact only weakly with the detectors, the field is growing exponentially, and will eventually overcome the weakness of the coupling. 
The highly mixed state of the field therefore leads to a highly mixed state of the detectors.
Beyond the number of iterations set by Eq.~\ref{instability}, we expect that the state of the detectors will become sufficiently mixed that no further  entanglement can be extracted.
However, if $\abs{d_1}$ is sufficiently close to $1$, the number of iterations is large and much entanglement can be extracted before the instability interferes with entanglement extraction.

We will see that if a cutoff is introduced on the field modes so that only the few modes closest to resonance are included, the system approaches a stable fixed point.
These modes are the most relevant to the entanglement extraction process, so this fixed point captures well the dynamics of the system up to a very large number of iterations.

The dynamics of the system is therefore roughly characterized by three phases. In the first phase, the field converges rapidly toward the approximate fixed point.
In this phase the energetic cost and extracted entanglement depend strongly on the initial state of the cavity field (we will be exploring this physics in the sections below). 
In the second phase, the field has reached an approximate fixed point, and the modes most relevant for entanglement extraction have lost their memory of the initial state. The state of the field in this resonant window no longer changes between cycles, and it is only the highly off-resonant modes that are being very slightly modified every cycle (specifically, becoming more energetic and mixed).
In the third phase, after many iterations, the off-resonant modes become highly excited enough that they begin to have a significant effect on the field's ability to entangle pairs of detectors. Once this happens the extracted entanglement per cycle correspondingly drops off, eventually reaching zero.
The time scale for this instability is only weakly dependent on the initial state, and can be estimated from the dynamics alone.

\subsection{Sustainable entanglement extraction}

One might expect intuitively as we proceed through detector-field interaction cycles that the field's ability to impart entanglement to the detectors would decrease. 
This is because of noise generated by switching on and off the interaction during each cycle; since the cavity is of a finite volume these excitations reflect off of the walls of the cavity rather than propagating away.
Furthermore, any correlations produced between the detectors and the field each cycle would be expected to induce additional noise in the field as we remove each pair.
Both of these effects are expected to increase the mixedness of the state and therefore to extinguish the entanglement that can be acquired by putting two detectors into interaction with it. 
Thus, if a fixed point of the field exists one might expect it to be unable to impart entanglement to the detectors via additional interaction cycles.

\begin{figure*}[t]
	\centering
                 \includegraphics[width=0.46\textwidth]{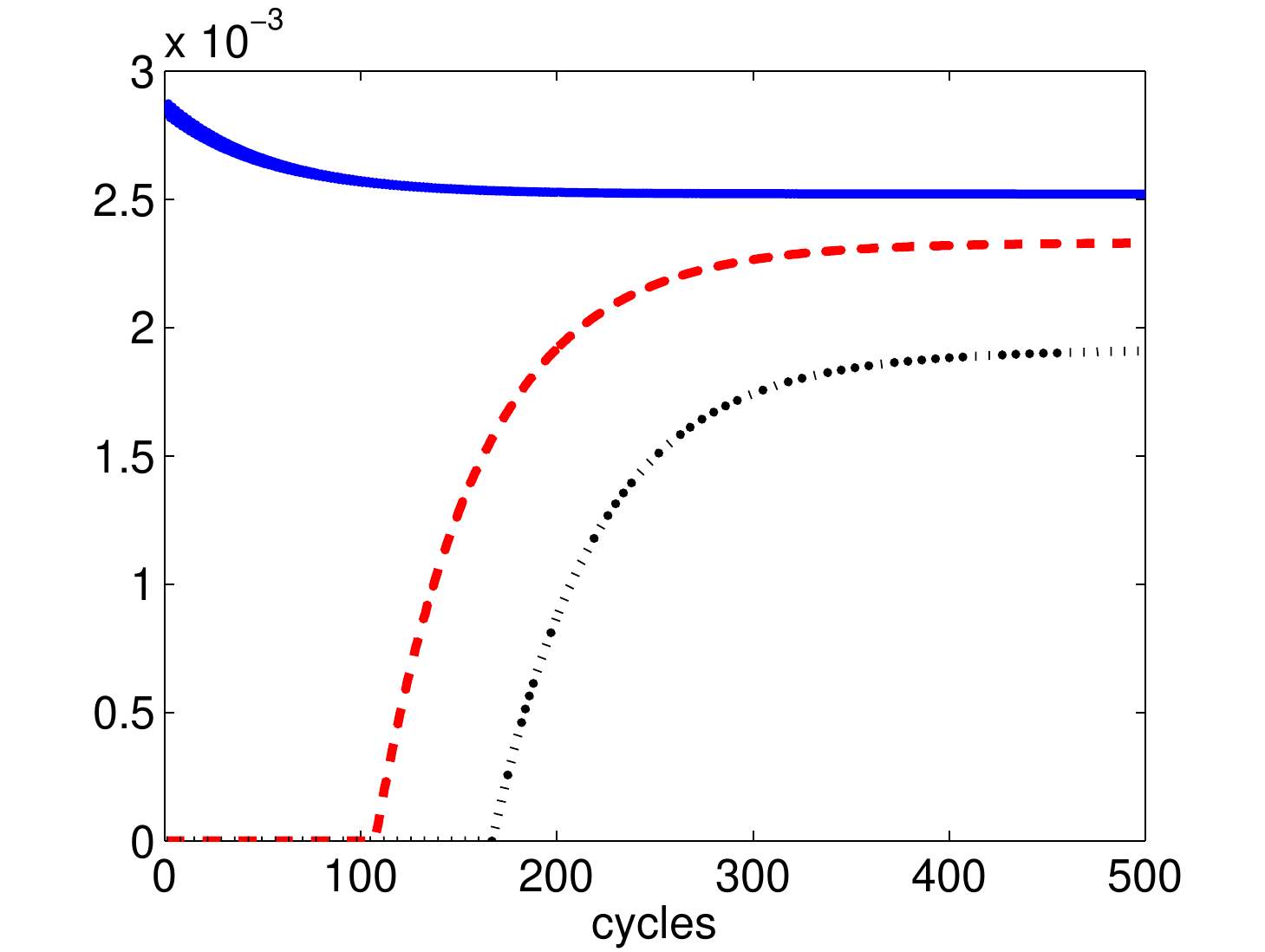}
                 \includegraphics[width=0.46\textwidth]{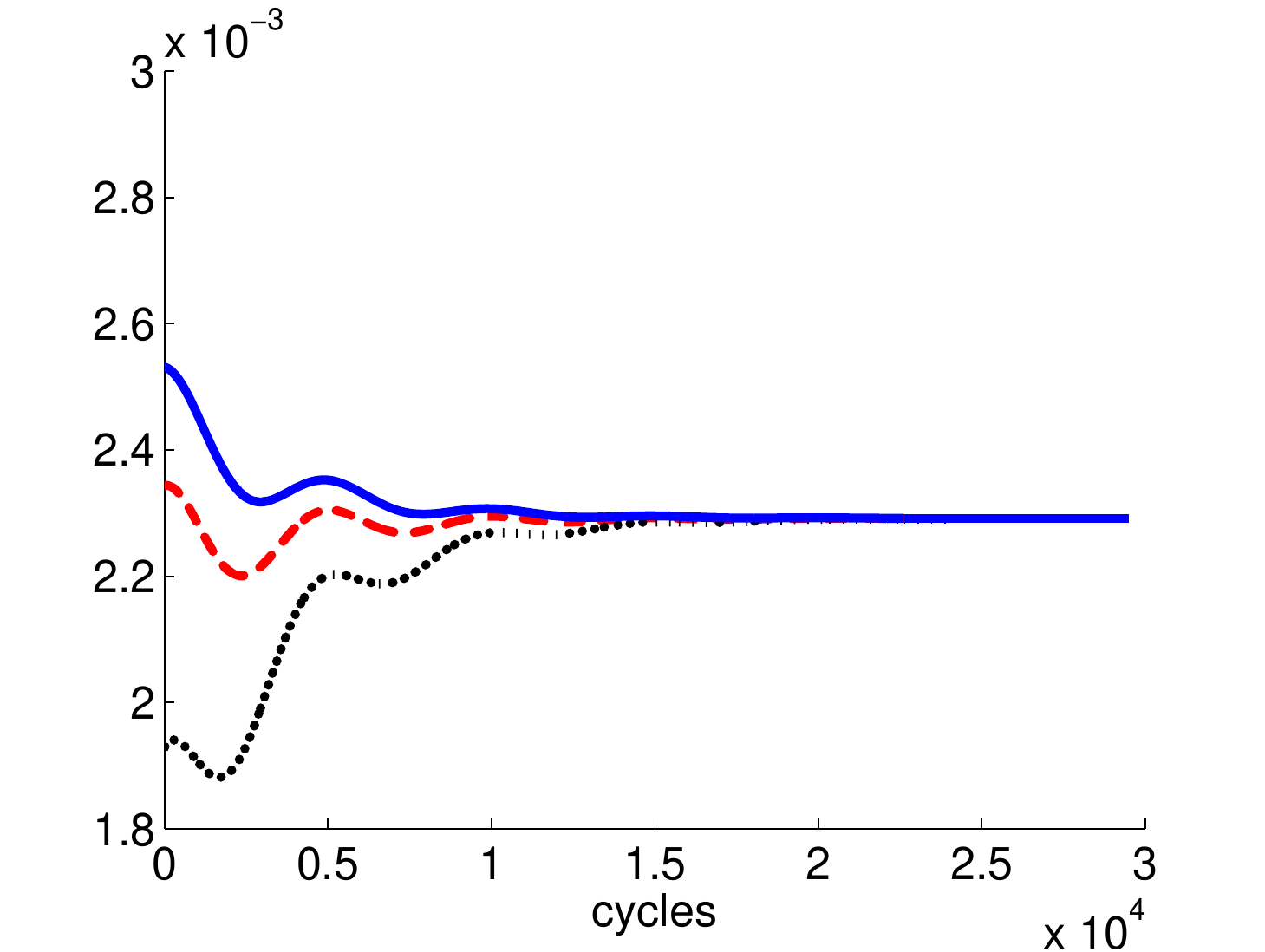}
	\caption{We plot the logarithmic negativity attained per cycle by detector pairs, as a function of cycle number. On the left this is plotted for a small number of cycles, and on the right for a large number. The three lines correspond to different initial states of the field. The solid (blue) line corresponds to the vacuum state, whereas the dashed (red) line corresponds to a thermal state of temperature $T=0.5$ and the dotted (black) line to a temperature of $T=1$. The time of evolution per cycle was set to $t_f=20$.}
        \label{lognegplot}
\end{figure*}

However, we have found this intuition to be incorrect.
Under a suitable mode truncation, the field does indeed approach a fixed point, 
and this fixed point does in fact entangle fresh detectors every cycle, even though the field itself is unchanged between the beginning and end of each such cycle. This can be confirmed by directly computing the fixed point via Eq. \eqref{fixedsol}. We must question, however, attractiveness and stability of this state. We find that the fixed point is in fact very stable and attractive, and indeed independent of the initial state of the field. Remarkably,  this means that even if the initial state of the field is such that entanglement cannot be extracted during a single cycle, the repetition of the process drives the field to a state for which it can. To quantify the amount of entanglement extracted we have used the aforementioned logarithmic negativity, which has the nice feature of being additive under tensor product \cite{plenio}. 

Plotted in Fig. (\ref{lognegplot}) is the logarithmic negativity imparted per cycle onto the detector pairs, as a function of cycle number. This is for the case that the time of evolution is larger that the light-crossing time between the detectors, specifically with $t_f=20$. This was performed using the algorithm laid out in Sec. \ref{modelSect}. We include two plots, one examining the short term behavior (i.e. small number of cycles) as well as that of the long term (large number of cycles, but not large enough to observe the decohering effect of the highly off-resonant modes). In each plot we display the results for several different initial states of the field, specifically the vacuum state and two thermal states. In the short term regime we observe that, as expected, thermal fluctuations in the field prohibit the extraction of entanglement. After several cycles, however, the field has been driven to a non-thermal state (see subsection \ref{thermalSect}) such that extraction becomes possible. In the long term behavior we observe two important points. Firstly, after many cycles the amount of entanglement obtained per cycle becomes constant and non-zero. At this point we have effectively reached the fixed point, and the state of the field is then observed to match the state obtained directly via Eq. (\ref{fixedsol}). As claimed, the entanglement obtained every cycle is non-zero, even though the field is no longer changing. Second, it is clearly seen that the fixed point behavior is independent of what state the field was initiated in. It can be confirmed that the field state in all three cases reaches the same fixed point. In subsection. \ref{thermalSect} we further discuss several important aspects of this evolution in regards to initial thermality of the field.

Summarizing this result: a cavity with an arbitrary state of the field can, in theory, be used as a source of entanglement for an extremely large number of detector pairs, independently of the initial state of the cavity. This knowledge can be useful in order to devise experimental implementations where the quantum field is used as a renewable entangling resource. 

%

Of course, there is an energy cost to this process. This cost is easily computed by taking the difference of the expected energies of our system (detectors and field) at the end and beginning of each cycle, where the energy is computed via Eq. (\ref{energy}). This would then be the energy per cycle that must be input. For example one can compute the energy cost of the procedure once the fixed point has already been reached and find that it is of course finite and positive. Since the field is not changing, this is simply the energy gained by the pair of detectors over the course of a cycle.

\begin{figure}[t]
	\centering
        \includegraphics[width=0.45\textwidth]{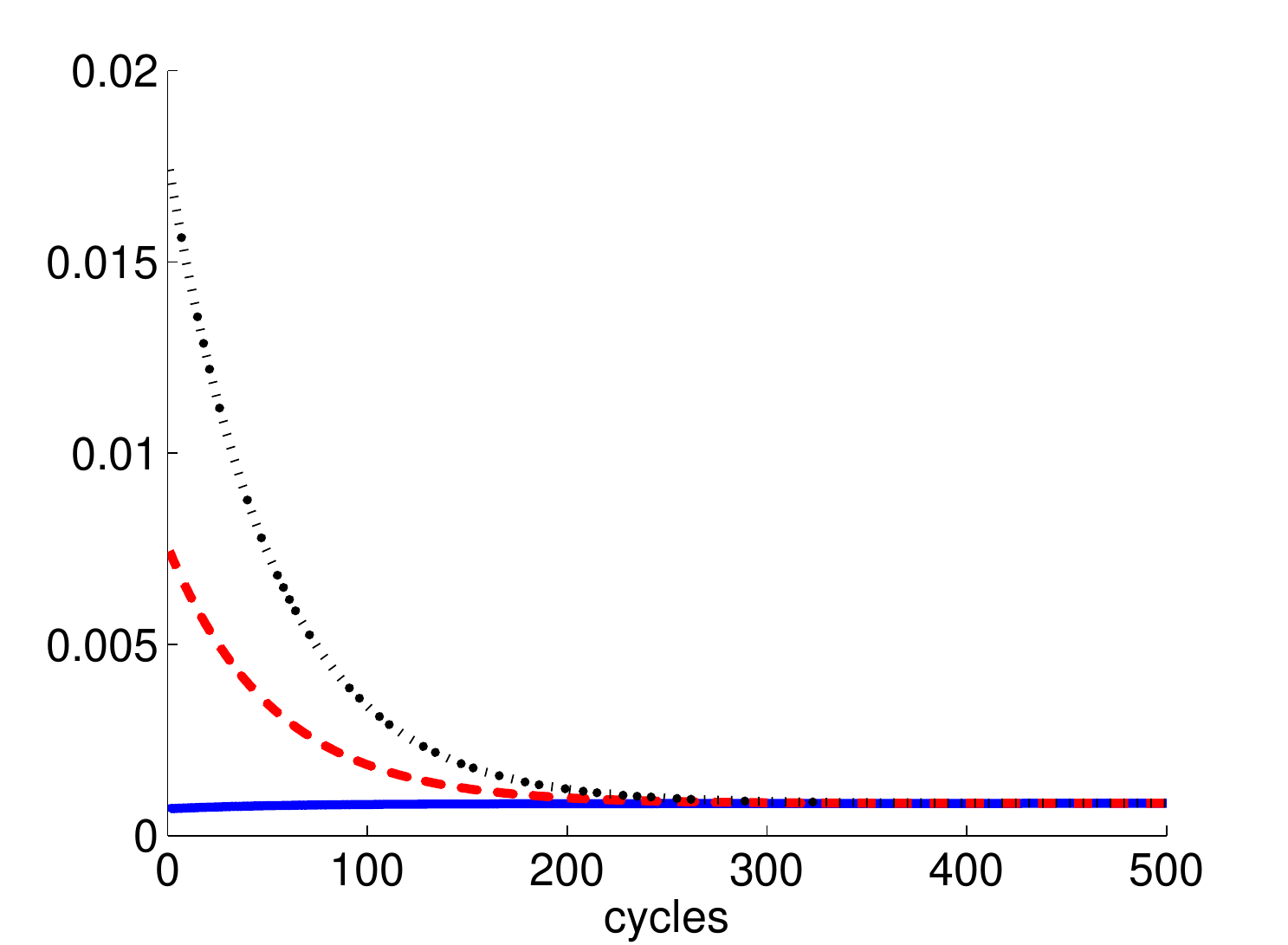}
	\caption{The energy input per cycle as a function of cycle number. The solid (blue) line corresponds to the case in which the field is initiated in its vacuum state, whereas the dashed (red) line corresponds to a thermal state of temperature $T=0.5$ and the dotted (black) line to a temperature of $T=1$. The time of evolution per cycle was set to $t_f=20$.}
        \label{energyfig}
\end{figure}

As an interesting aside, we have computed the energy input per cycle in the short time regime described above. The results are displayed in Fig. (\ref{energyfig}). We see that, interestingly, thermality of the field greatly increases the energy required to drive the field to a non-thermal state that allows entanglement extraction.

\subsection{Initial thermal field}   \label{thermalSect}

In the previous subsection, as well as in this section, the number of cycles that we have considered is small enough that only the near-resonant modes of the field are relevant for the behavior of the detectors. The instability associated with off-resonant modes (which will be discussed further in the next subsection) occurs only after many more cycles have occurred. In this subsection we will be examining certain properties of the field during the first few cycles. In doing so we only include in this analysis the modes that are relevant for the detectors, such that we can easily identify the important aspects of the field's evolution over several cycles with the evolution of the detector-detector entanglement seen in the previous subsection.

When allowing long interaction times per cycle we have seen that we can harvest entanglement from a cavity field without modifying the state of the field, thus proving its sustainability. A question that arises, however, is how easy it would be to prepare the field in this specific fixed point. As we have seen in the right-hand plot of Fig. (\ref{lognegplot}) this turns out to not be a problem. The fixed point is stable and attractive, and is reached independently of the initial state of the field.

We also see in the left plot of this figure that, as expected, thermality in the initial field state interferes with our ability to extract entanglement from it or to acquire entanglement via effective interaction of the detectors through the field. With increasing temperature, the field quickly becomes incapable of providing any entanglement at all. If, however, we start cycling detectors through the field, a regime will be reached at which the state of the field has been modified enough that we can begin to obtain entanglement, before it continues on to eventually converge to the fixed point. We have also observed that this driving of the field towards an entanglement-enabling state requires significant amounts of energy as compared to the energy expense once the fixed point is reached. Here we study further this initial transition that is observed when the field is initialized in a thermal state.

\begin{figure}[t]
	\centering
        \includegraphics[width=0.45\textwidth]{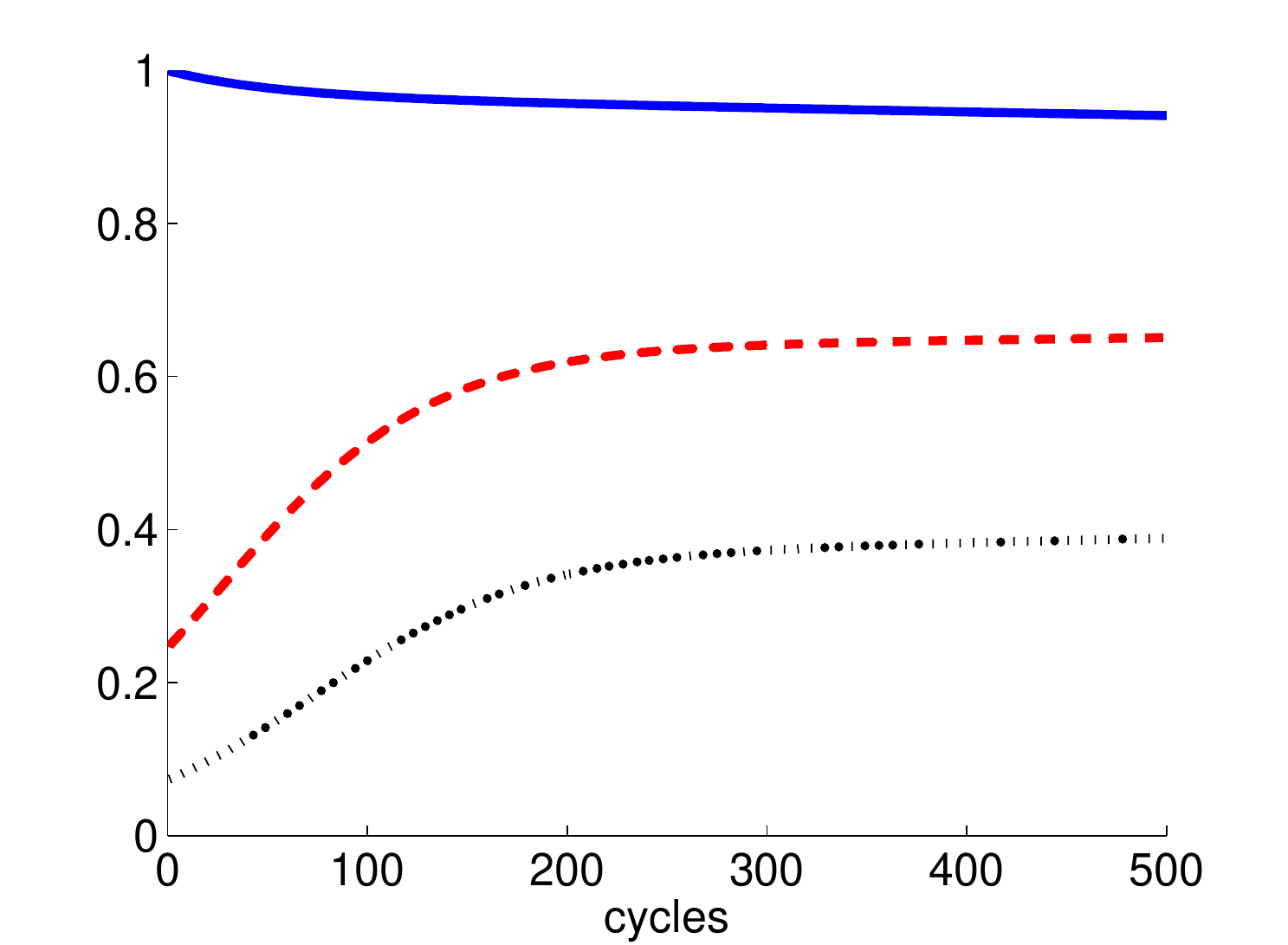}
	\caption{The field purity as a function of cycle number. The solid (blue) line corresponds to the case in which the field is initiated in its vacuum state, whereas the dashed (red) line corresponds to a thermal state of temperature $T=0.5$ and the dotted (black) line to a temperature of $T=1$. The time of evolution per cycle was set to $t_f=20$.}
        \label{thermPure}
\end{figure}

During this transition period the field is being driven to a more pure state. This can easily be seen using the measure of purity Eq. (\ref{purity}). Plotted in Fig. (\ref{thermPure}) is the purity of the field under the same evolution. Thermality of the field obviously means that the state's purity is initially quite low, but we observe that over the course of several cycles the detectors are acting to purify the field during their time in the cavity.


A naive interpretation of this phenomenon would be based on the observation that we are continually inserting cold oscillators (they are in their ground state) into a hot cavity. This would seem to indicate that the freshly injected oscillators are acting to cool down the field until the temperature is low enough to allow entanglement extraction. 

We have found that that this interpretation is incorrect, however, because during the period of decreasing mixedness the field is also becoming very non-thermal. This can be seen using the relative entropy as a measure of the distance from thermality, as discussed in Sec. \ref{sectRel}. With respect to this measure the closest thermal state to a given state $\rho$ is that which has the same energy $E_\rho$ as $\rho$. The relative entropy between $\rho$ and this closest thermal state is then the difference of their entropies: $S_\text{th}-S(\rho)$, where $S_\text{th}$ is the entropy of the closest thermal state. We must be careful when using this however because for low energies (corresponding to low temperatures) the smallness of $S_\text{th}$ may lead us to believe that $\rho$ is very nearly thermal, even if it is as far from thermality as can be achieved while staying on the energy shell $E_\rho$. To counter this we instead use a relative distance measure by dividing by $S_\text{th}$. Lastly, since we would like a ``thermality estimator", we will impose that it equates to unity in the case that $\rho$ is exactly thermal. With this, we define our thermality estimator $D$ as
\begin{align}
	D(\rho) \equiv 1-\frac{S_\text{th}-S(\rho)}{S_\text{th}}=\frac{S(\rho)}{S_\text{th}}.
\end{align}
This measure is bounded from above by one, which is saturated when $\rho$ is exactly thermal, and is bounded from below by zero, which is achieved in the case that $\rho$ is pure.

\begin{figure}[t]
	\centering
        \includegraphics[width=0.45\textwidth]{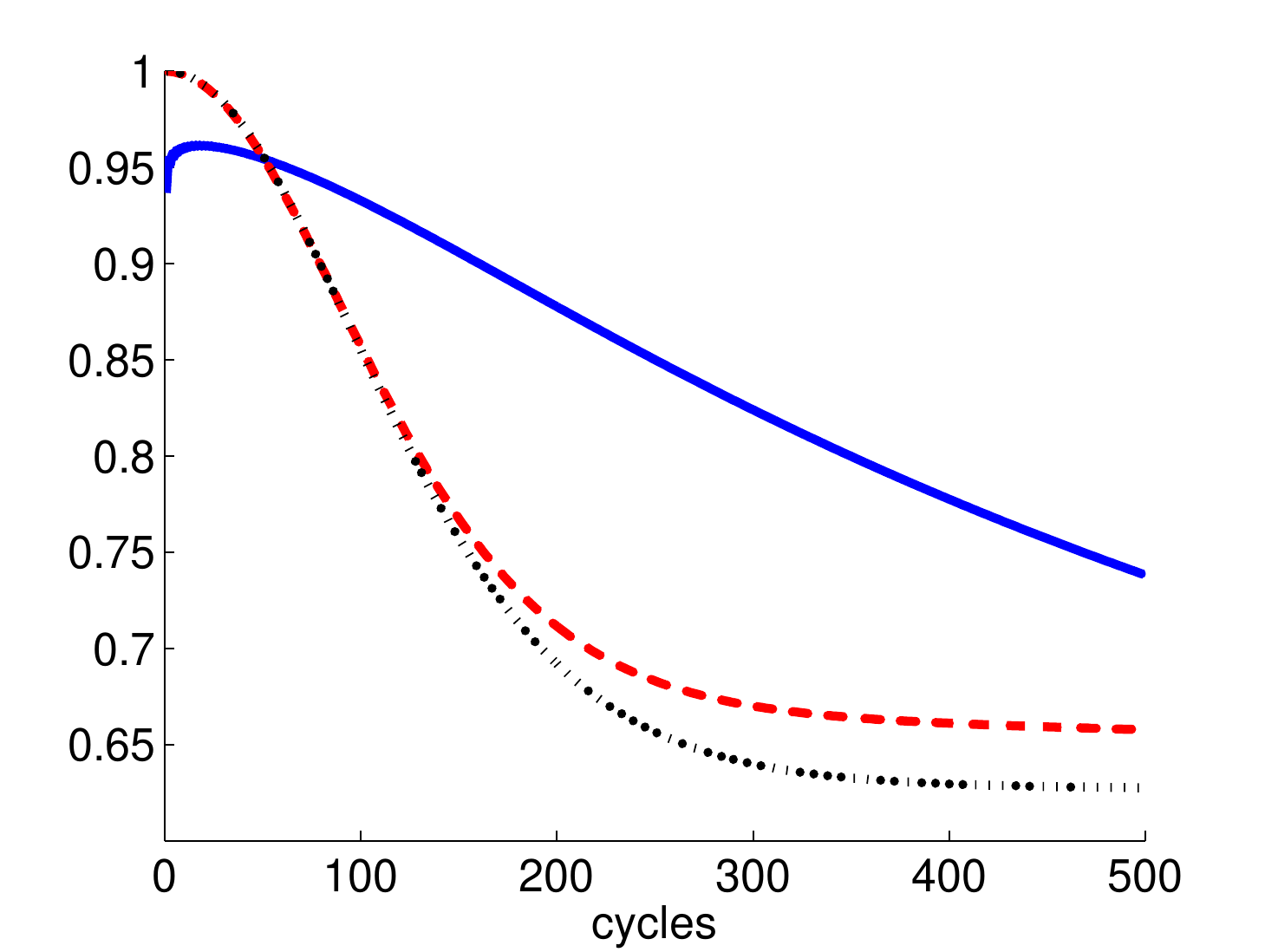}
	\caption{The thermality estimator $D$ as a function of cycle number. The solid (blue) line corresponds to the case in which the field is initiated in its vacuum state, whereas the dashed (red) line corresponds to a thermal state of temperature $T=0.5$ and the dotted (black) line to a temperature of $T=1$. The time of evolution per cycle was set to $t_f=20$.}
        \label{thermality}
\end{figure}

We are now able to plot $D(\rho)$, applied to the field, as a function of the cycle number and observe that, as stated, the field becomes very non-thermal during the time that it is becoming less mixed. This is plotted in Fig. (\ref{thermality}). Note that the odd initial value of $D(\rho)$ observed when $\rho$ is the vacuum state is due to the fact that $D(\rho)$ is ill-defined in the unique case of the vacuum state. In the plot the initial point actually represents the field after one cycle has been performed, in order to avoid this problem.

This shows that we are not trivially cooling down the field. Rather, the field is being driven far off the thermal manifold into a non-thermal, entanglement providing state. In fact, for the parameters being used here it is easy to check for the case of initial temperature $T=1$ that when considering the field state after $500$ cycles (i.e. at the ends of the figures) the closest thermal state (the one with the same energy as the field) does not provide any entanglement to the detectors. Additionally if we instead consider the thermal state that has the same mixedness as the field, this state also provides no entanglement. This is despite the fact that the actual state itself \emph{does} provides entanglement, as seen in Fig. (\ref{lognegplot}). Remarkably, the field appears to be non-thermally driven to a state that is very well designed for providing entanglement, and this is far from a simple cooling process.

As an aside, it is interesting to note that if we increase the time of evolution per cycle then the field is actually driven to an even less thermal state. This is contrary to the intuition that a longer interaction-time might mean that the field thermalizes with the detectors, and thus that it would remain or become relatively thermal as we cycle through detector pairs. This reinforces the idea that the process of cyclicly repeating the entanglement extraction protocol cannot be understood as a cooling process.

\subsection{Onset of the instability}  \label{Sectinstability}

We discussed above that we can find a fixed point in the evolution provided that we truncate the mode expansion of the quantum field.
This truncation accurately captures the relevant dynamics of the protocol even for a very large numbers of iterations.
However, when all the higher, off-resonant modes are included we generically find that the cavity state eventually becomes unstable (although only  after an extremely long number of cycles), and therefore that the mode truncation is not reliable past a given number of iterations of the protocol.

What occurs when highly off-resonant modes are included is as follows. In each individual cycle, the effect that the oscillators have on these off-resonant modes is minute, as it must be, and similarly the effect that these modes have on the detectors is insignificant (assuming the modes are not yet extremely energetic). Over many cycles however, the tiny effects appear to accumulate such that eventually the off-resonant modes become highly excited and mixed enough that they begin to have an observable effect on the detectors.  At some point, the detectors become sufficiently entangled with the off-resonant modes that tracing over these modes when the detectors exit the cavity has a sufficiently decohering effect that the entanglement generated between the detectors in each cycle drops off to zero. 

\begin{figure}[t]
	\centering
        \includegraphics[width=0.45\textwidth]{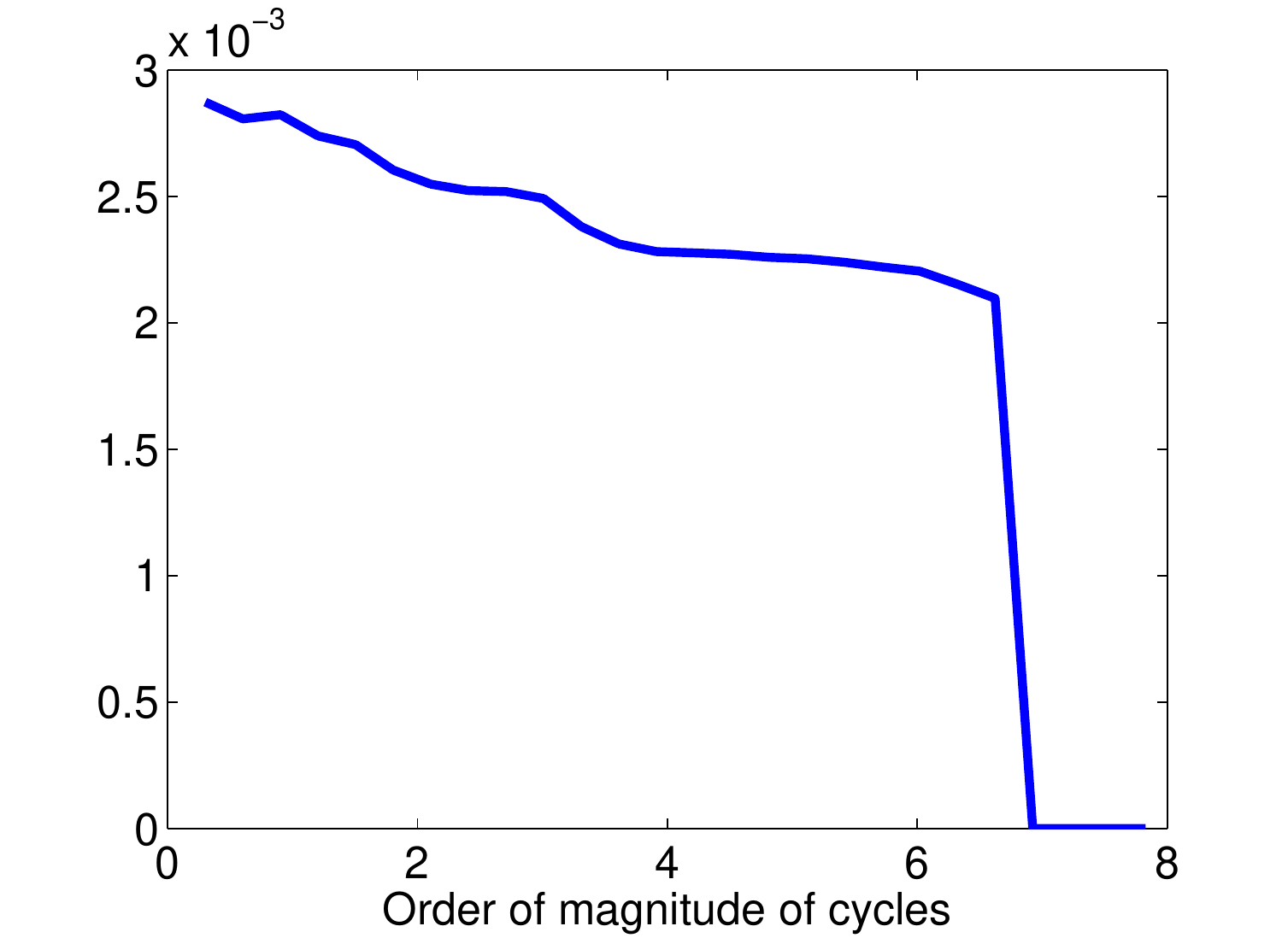}
	\caption{The logarithmic negativity obtained by the detectors as a function of $\log_{10}(\text{number of cycles})$. The field was initiated in its vacuum state. We see that after a large number of cycles, approximately $10^{7}$, the entanglement decays to zero as a result of the increasingly excited off-resonant modes. Note that the sharpness of the drop is due to a lack of resolution, since this plot was generated by evaluating the extracted entanglement at every $2^n$'th cycle.}
        \label{ultralong}
\end{figure}

We demonstrate this effect in Fig. (\ref{ultralong}), where we plot the logarithmic negativity as a function of the number of cycles on a logarithmic scale. The field was initiated in its vacuum state, and all parameters in this plot are the same as those in Fig. (\ref{lognegplot}).
Note that we can see the same plateau of about $E_N \approx 2.3 \times 10^{-3}$ as observed in Fig. (\ref{lognegplot}), but we also discover now that over the course of many more cycles the generated entanglement begins to slowly degrade as the off-resonant modes become more highly excited before eventually dropping off to zero. It should be noted that even though the total amount of extractable entanglement is limited when the highly off-resonant modes are included, the total amount of logarithmic negativity (added over all detector pairs) that could be obtained from the scenario in Fig. (\ref{ultralong}) is still on the order of $10^{4}$ which is enough distillable entanglement to produce a large number of Bell pairs via entanglement distillation.  Note that the sharpness of the drop in this plot is due to a lack of resolution, since it was generated by evaluating the extracted entanglement at every $2^n$'th cycle.

As we showed in section \ref{fixed}, the stability of the system is governed by the maximal eigenvalue of the field superoperator. The maximum eigenvalues are expected to derive from the off-resonant modes, and thus by Eq. (\ref{instability}) we are able to estimate the number of cycles $n$ needed for the off-resonant modes to make significant contributions to the detectors, and thus to interfere with entanglement generation. We have observed that indeed this method gives a good estimate of after how many cycles the entanglement drops to zero, and we can thus use this eigenvalue approach to easily explore the behavior of the critical number of cycles $n$ at which further entanglement extraction becomes impossible.

\begin{figure}[t]
	\centering
        \includegraphics[width=0.45\textwidth]{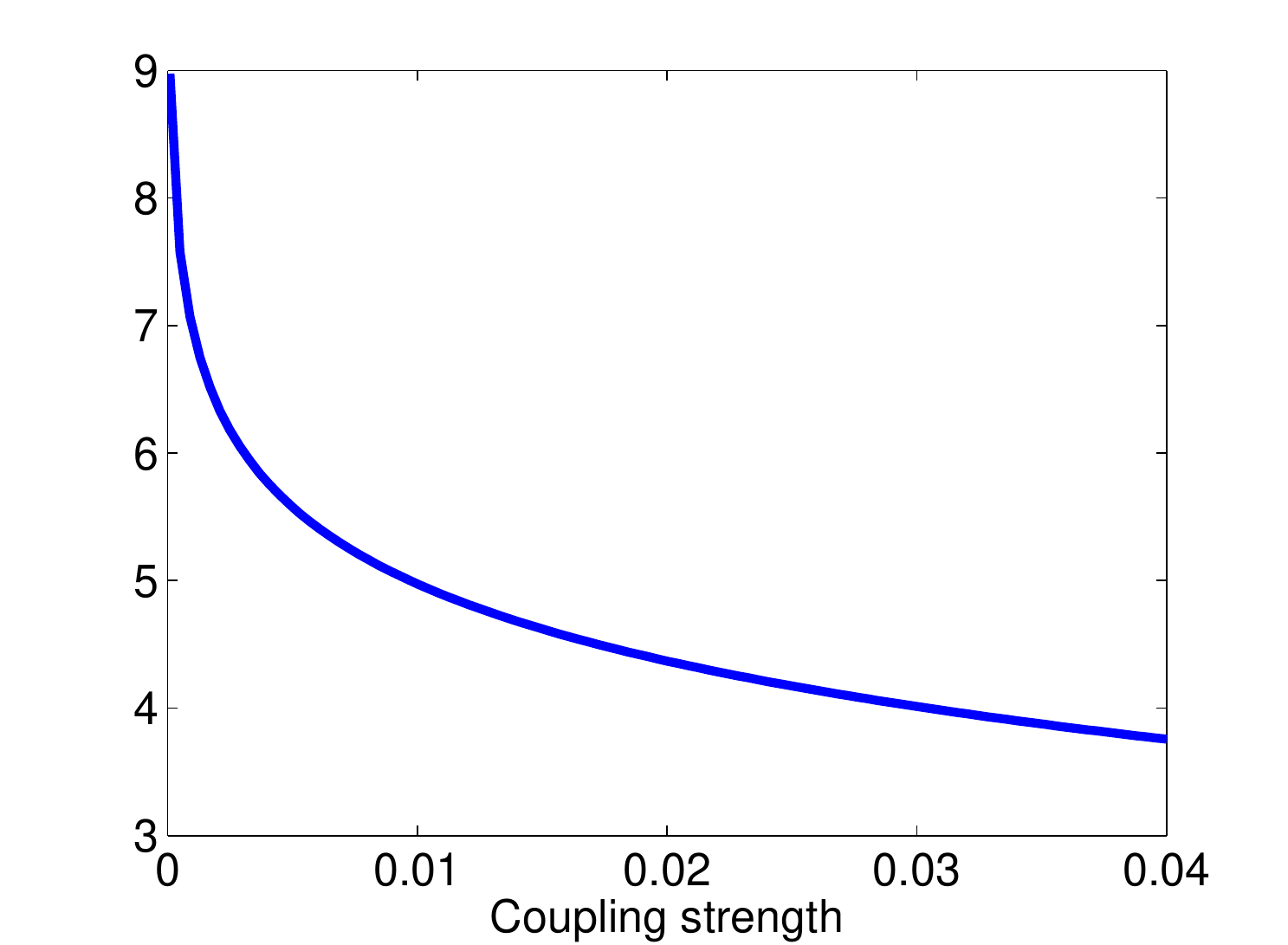}
	\caption{The order of magnitude of the number of cycles needed for entanglement extinction as a function of the coupling strength $\lambda$.}
        \label{eigcoupling}
\end{figure}

For example, we plot in Fig. (\ref{eigcoupling}) the critical number of cycles (on a logarithmic scale) as a function of the coupling constant $\lambda$. Here the evolution time per cycle was set to $t_f=21$. We see that increasing the coupling reduces the critical number of cycles. This makes sense physically as a larger coupling results in a larger impact per cycle on the off-resonant modes, meaning that they don't need as many cycles to become excited enough to the point of relevance.
This behaviour can also be understood mathematically as a consequence of the weakness of the coupling between the detector and quantum field.
To zeroth order in $\lambda$, the eigenvalues of the matrix $\mat{D}$ are pure phases, so have unit absolute value.
The first perturbative corrections to $\mat{D}$ occur at second order in $\lambda$, so we expect the maximal eigenvalue to be approximately $1 + O(\lambda^2)$. Translating this into $n$ via Eq. (\ref{instability}), it is consistent with the numerical results. 

\begin{figure}[t]
	\centering
        \includegraphics[width=0.45\textwidth]{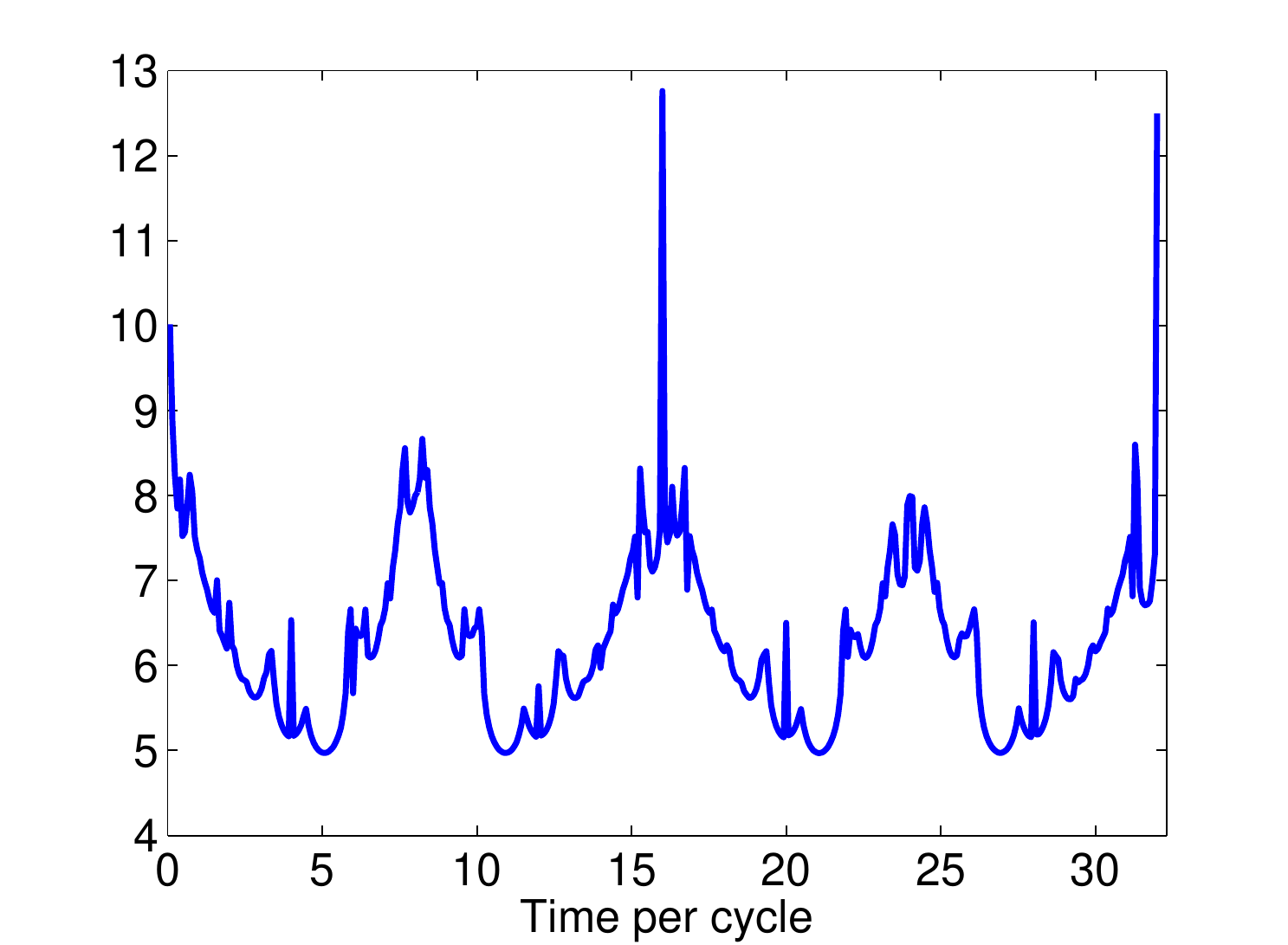}
	\caption{The order of magnitude of the number of cycles needed for entanglement extinction as a function of the cycle time $t_f$. This magnitude is periodic with a period of $16$, which is twice the cavity length.}
        \label{eigtime}
\end{figure}

We also plot in Fig. (\ref{eigtime}) the critical number of cycles (on a logarithmic scale) as a function of the time per cycle $t_f$.
We would like to remark and stress this result, because when the time evolution per cycle is large enough it is commonplace in the field of cavity quantum optics to resort to the so-called single mode approximation \cite{ScullyBook}. This approximation consists of neglecting the dynamics coming from the coupling of the off-resonant modes with the particle detectors. 
It is easily seen that the effect of the off-resonant modes in the quantum state of a particle detector at leading order in perturbation theory becomes negligible for times much larger than the light-crossing time of the cavity (see for instance \cite{Robert2013}). 

Importantly for our scheme of cyclic entanglement extraction, the critical number of cycles as a function of the cycle duration is not monotonically increasing with $t_f$ as might naively be expected from the single mode approximation.
In this case it is important to remember that the single mode approximation is effective because one can neglect the contribution of the off-resonant modes \emph{relative to the resonant modes}.
However, it is not the case that the contribution of the off-resonant mode decays with the time per cycle; rather, it is the contribution of the resonant modes that is growing.
The relevant coupling of the off-resonant modes is oscillatory in time rather than decaying, as is captured in Fig.~\eqref{eigtime}.
Thus, while we may be well within the domain of applicability of the single-mode approximation for describing a single iteration of our protocol, this approximation is not appropriate for describing the dynamics after a very large number of iterations.


Since the instability is an effect of the off-resonant modes, it is sensitive to the
switching function of the detectors.
Here we have assumed sudden switching, for which the strength of the interaction with a given mode decays quadratically with the mode's frequency.
For smoother switching functions the decay may be of a higher power, exponential, or even Gaussian.
If the switching is done in a smother way one would therefore expect the time scale of the instability to increase.
The role of the switching function is deserving of further study, but this is left as an area for future work.

We conclude this section with an intuitive  way to understand how the off-resonant modes are destabilizing the fixed point: The system has a parameter that describes the coupling strength, namely the parameter that we use to switch the interactions on and off. This means that we are dealing with a system that is parametrically driven, with the possibility of parametric resonance. All the field modes are parametrically driven by the rhythm of switching on and off the interaction, i.e., by the entering and exiting of the detector pairs. The cycle length (and the form of the switching function) is important because it determines which frequencies the switching will parametrically excite. The Fourier transform of the switching function (and thus the cycle length and smoothness of the switching) determines which modes are by how much parametrically excited. We postulate that as the off-resonant modes become parametrically driven they are exponentially fast excited, as is common for parametric resonance, up until at some point where their population is so large that non-perturbative effects destabilize the fixed point. For the destabilization to happen faster, some off-resonant modes' frequencies should therefore be represented strongly in the Fourier transform of the switching function. This may account for the nontrivial dependence of the destabilization time as a function of the cycle length, as shown in Fig.~\eqref{eigtime}. This perspective will also be explored in later work.

\subsection{Is a fixed point possible for short cycle times?}

We discussed above that there is always a metastable fixed point
provided that the duration of every cycle is large enough, meaning that up until a great many cycles have occurred only a small number of modes are relevant to the dynamics of the detectors.
However, we have found that there are some regimes where the application
of the cycle repetition protocol does not lead to a fixed
point. In other words, when the duration of the cycle $t_f$ is
below some threshold the system does not reach a fixed
point. While this minimum duration does depend on the
particular parameters of our setting, the cycle time scales
where there is not a fixed point are of the order of the
light-crossing time between the two detectors (and therefore in the zone where spacelike entanglement harvesting occurs  \cite{Reznik1}).

\begin{figure}[t]
	\centering
        \includegraphics[width=0.45\textwidth]{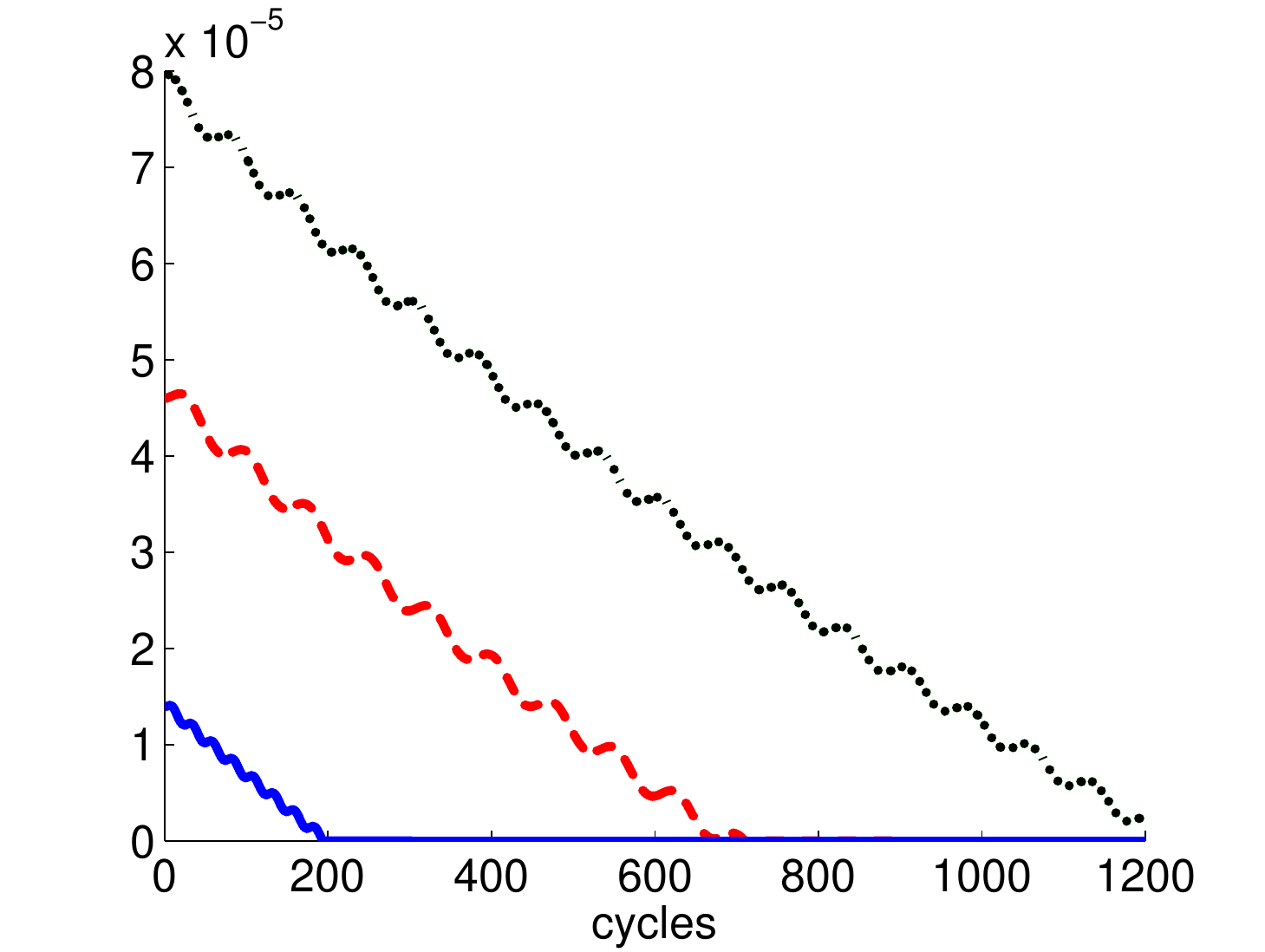}
	\caption{The logarithmic negativity obtained per cycle in some regimes where there is no fixed point. The field was initiated in its vacuum state. The three lines correspond to three different interaction times per cycle, $t_f$. We represent this in terms of the light-crossing time between the detectors, which since the speed of light is set to unity is just the distance $r=8/3$ between them. The solid (blue), dashed (red), and dotted (black) lines correspond to $t_f=1.44r$, $t_f=1.48r$, and $t_f=1.52r$, respectively.}
        \label{extinction}
\end{figure}

An interesting question to ask now is what happens with our ability to entangle the two detectors in the ground state by coupling them to the field in these regimes where a fixed point does not exist. In other words, can we harvest a considerable amount of entanglement from the field in the regime where the detectors remain spacelike separated with the technique presented here? We can answer that question by evaluating the amount of entanglement extracted per cycle when the cycle time is below the fixed point threshold. Note that in doing this, due to the smaller time of evolution, there are many more modes that are relevant to the dynamics of the detectors and therefore must be included in the numerical analysis of the problem. We plot in Fig. (\ref{extinction}) the logarithmic negativity obtained per cycle using much shorter interaction cycles than that considered above (although still slightly greater than the light-crossing time). We see that in this regime we cannot reliably extract entanglement over many cycles, and in fact over the course of many cycles the field is driven towards a less-entangling state (as initial intuition would suggest) rather than the opposite behaviour observed above for the long cycle duration regime.

This provides a convenient opportunity to give some further discussion on whether or not a fixed-point field is possible in the $t_f<r$ regime (spacelike separation), and whether or not such a fixed point endows entanglement onto the detectors. Numerically, and at least in the current scenario, the answer appears to be that a fixed point is not reached. We will discuss that while there are some obvious arguments to make towards why this must necessarily be the case based on intuition coming from information theory, we will argue that such an argument is flawed in the case of quantum fields and there is actually no reason to think \emph{a priori} that such a fixed point is necessarily impossible.

The naive argument for why it must be impossible goes as follows. Assume there exists a fixed point in the spacelike regime ($t<r$) which provides entanglement to the detector pairs. The pair of detectors never come into causal contact, so that can only interact locally with the field. But entanglement cannot be increased under local operations. Therefore, the total entanglement of the detectors plus that of the spacelike separated regions of the field at their locations can not be increased, and thus if the detectors become entangled this entanglement should have previously resided in the field. Therefore, the appearance of entanglement in the detectors should impact the state of the field and make the existence of a fixed-point impossible.

Even if the detectors do not become entangled one can make a similar argument using any correlations in general (for example as quantified by the mutual information). A detector-field interaction of the type considered, giving an evolution of the form in Eq. (\ref{soln}), will generically generate correlations between the detectors starting immediately at time $t=0$. To see this, one can Taylor expand the exponential in Eq. (\ref{soln}) with respect to $t$ and easily show that the off-diagonal block $\mat{\gamma}_{12}$ of the detector-detector covariance matrix generically grows as $t^2$ for small $t$. However, correlations in general cannot be increased solely by local interactions, and thus if the detectors never come into causal contact the same argument as above still applies: the correlations must be extracted from the field state and therefore the field state must necessarily change. This is because, even if a fixed-point field does not provide entanglement to the detectors, it will in general provide other correlations. Thus, we come to the conclusion that a fixed-point is impossible to obtain in the $t_f<r$ regime.

The problem with this argument is that it omits important aspects of our protocol. In particular, each detector can locally generate correlations between either itself and a field quanta or between multiple quanta, and the free evolution of the field can transport these correlations to different parts of the cavity.  For $t_f<r$ these correlated quanta of course cannot transport their newly produced correlation to both detectors over the course of a single cycle, but they can do so over multiple cycles. For example imagine that we have obtained a fixed point, such that the state of the field is the same at times $t=0$ and $t=t_f$, and an aspect of this state may be that it contains a travelling pair of correlated quanta that were produced during the previous cycle and will be absorbed by the detectors (one quanta for each) during the next cycle. That is, local field correlations generated in a previous cycle may then be transferred to the detectors at a later cycle. Another way to state this type of scenario is that the detectors indeed drain the field of correlations, but that the free evolution of the field is then allowed to regenerate this loss. 

Thus while we have described previous protocols as ``harvesting'' entanglement from a quantum field, here the protocol is analogous to farming entanglement: in addition to extracting entanglement from the quantum field, the interaction with the detectors is also ``sowing'' the seeds of entanglement to be extracted in subsequent cycles.
However we know from Ref.~\cite{Reznik1} that harvesting entanglement at spacelike separation is a more difficult process.
While we see no reason in principle that it cannot be done, we have not yet found detector settings that allow for entanglement farming in the spacelike regime.



\section{Summary and Outlook}

We studied a protocol for harvesting and even sustainably farming quantum entanglement from a quantum field in a cavity. The protocol consists of placing temporarily two `detectors', i.e., two localized quantum systems (such as atoms, molecules or ions) that are in their ground state, into an optical cavity. They interact with the cavity field for some cycle time, $t_f$. The now entangled detectors exit the cavity and a new pair of  detectors in their ground state enters the cavity for a new cycle to begin. 

The entanglement that the detectors acquire in each cycle has two distinct sources: On the one hand, the detectors can swap entanglement from the cavity field, a process that can take place even when the detectors remain spacelike separated while in the cavity. We have found this type of entanglement harvesting to be transient, however, i.e., we have not found a regime in which this entanglement extraction could be sustained over many cycles. It should be interesting in this context to study if there exists a fundamental limit to entanglement extraction in this case, perhaps similar to the fundamental limit to work extraction in a Carnot cycle. 

On the other hand, the detectors can become entangled by interacting via the cavity field if the cycle time is long enough to allow the causal exchange of quanta between the detectors. 
To this end, we first studied the case where the highly off-resonant modes of the cavity can be neglected and where the cycle times are long enough to enable causal detector interactions via the cavity field. In this case, after some number of cycles the cavity field reaches a fixed point state that is highly non-thermal and independent of the field's initial state. Perhaps surprisingly, this fixed point state of the cavity allows for sustainable significant entanglement harvesting by the successive pairs of detectors. We determined the considerable amount of entanglement that can be harvested in this way, and computed the associated energy cost per cycle. 

Due to the sustainability of the method, we call this entanglement farming rather than mere entanglement harvesting. We also found that the field reaches a fixed point regardless of the initial field state in the cavity. This is of interest in regards to possible experimental realizations because it means that this useful fixed point state will be reached independently of any noise or imperfections in the preparation of the initial state of the cavity. The cycling of detector pairs in and out of the cavity drives the field to a steady-state that can be used to entangle detectors, even if the initial state of the field did not yield any entanglement.
For example, we have shown that it is possible to apply the protocol even to thermal states.

Our methods have also allowed us to examine a result that involves going far beyond the usual single or few mode approximations that are commonly used (see e.g., the Jaynes-Cummings model in \cite{ScullyBook}, or  \cite{Robert2013}). Namely, we were able to calculate the behaviour of the off-resonant modes for large cycle numbers. We found that there is a third stage of evolution after a very large number of iterations. In this final stage, off-resonant modes of the cavity become sufficiently excited as to make further entanglement harvesting from the cavity impossible. In practice, if the excitations of these off-resonant modes are sufficiently leaking, the third stage need not occur, so that the entanglement farming is sustainable indefinitely. 

In regards to intuition, our results indicate that the protocol of repeatedly inserting pairs of cold detectors in the cavity does not constitute an efficient cooling process, because the result is not a thermal state. Instead, our results indicate that the detector insertion and removals may be better understood as a periodic parametric driving, a perspective that will be pursued further in forthcoming work.

Additionally, we can view our setup as an entangling quantum gate which has the advantage that it is stably produceable and controllable by iterating a process that yields a fixed point. This combined with 1-qubit universal gates over atoms using relativistic motion (as in the scheme  proposed in \cite{AasenPRL}) would give a complete set of universal gates based on atomic motion in optical cavities that would allow for quantum computing out of relativistic effects. The main disadvantage of a scheme based on the setting presented in this paper would be that the two-qubit gates are also depurifying. It is expectable that variations in the interaction Hamiltoinan could lead to a fixed point with a smaller depurifying effect over the atoms.  I.e. We would like entanglement to be generated between the two atoms without increasing the entropy of the field so that it does not get entangled with the two atoms system but merely act as a carrier of their interaction. This path will be explored in future work.

Let us also discuss scenarios in which the setting presented here could be implemented experimentally. 
The most straightforward experimental implementation appears in the context of quantum optics. 
Namely, the theoretical scenario analyzed here could be implemented with optical cavities \cite{ScullyBook} being traversed by successive pairs of atoms along the transverse direction of the cavity.
The transverse profile of the modes effectively couples and decouples the interaction in a way similar to that suggested in this paper. 
The experimental setup could be similar to the kind used in Ramsey interferometry for quantum non-demolition measurement of the state of light in an optical cavity \cite{singlephotons}. 
Any such implementation requires of course a careful analysis of the cavity losses. 
But as discussed above, in a realistic experimental setup the higher frequency modes of the cavity have the lowest quality factor, and an enhanced leakage of higher frequency modes could be beneficial. This is because any such leakage would suppress the accumulation of field excitations in the off-resonant modes, which would in turn further stabilize the fixed point.    

Additionally, experimental implementations in other systems such as trapped ions or superconducting circuits appear to be within reach. Indeed, the kind of interaction Hamiltonians that we consider in this work can be implemented straightforwardly and with a high degree of control in both superconducting circuits and trapped ion settings (see, e.g., \cite{Diegger}).

We conclude with the observation that reliable mass production of entanglement, or entanglement farming, on the basis of a fixed point state should be possible also in many other experimental settings. Namely, instead of successively temporarily coupling pairs of particles to a cavity field, one may successively temporarily couple pairs or triplets, etc. of qudits to a suitable reservoir system. The qudits and the reservoir system could have any arbitrary physical realization, even outside quantum optics. In each cycle, $N$ fresh qudits in their ground states are coupled to the reservoir system, then removed.  The key requirement for the farming of $N$-partite entanglement by this method would be that the coupling between the qudits and the reservoir is such that the iterated coupling of fresh qudits drives the reservoir system to a fixed point state that is entangling the qudits.  As we saw here, such a fixed point does exist in the case where the qudits and the reservoir system are composed of harmonic oscillators, and we also found that the fixed point is completely stable if the number of harmonic oscillators in the reservoir is small enough.  It should be very interesting to find the sufficient and necessary conditions on qudit-reservoir systems for such a useful fixed point state of the reservoir to exist. 
 
\section{Acknowledgements} 
 
The authors gratefully acknowledge support through the NSERC's Banting, Discovery and Canada Research Chairs Programs.
 
\bibliographystyle{utphys}
\bibliography{Wilrefs}

\end{document}